\documentclass[12pt]{article}
\usepackage{graphicx}
\usepackage{epstopdf}
\advance\textheight by \topskip
\begin{document}


\newcommand{\HPA}[1]{{\it Helv.\ Phys.\ Acta.\ }{\bf #1}}
\newcommand{\AP}[1]{{\it Ann.\ Phys.\ }{\bf #1}}
\newcommand{\be}{\begin{equation}}
\newcommand{\ee}{\end{equation}}
\newcommand{\br}{\begin{eqnarray}}
\newcommand{\er}{\end{eqnarray}}
\newcommand{\ba}{\begin{array}}
\newcommand{\ea}{\end{array}}
\newcommand{\bi}{\begin{itemize}}
\newcommand{\ei}{\end{itemize}}
\newcommand{\bn}{\begin{enumerate}}
\newcommand{\en}{\end{enumerate}}
\newcommand{\bc}{\begin{center}}
\newcommand{\ec}{\end{center}}
\newcommand{\ul}{\underline}
\newcommand{\ol}{\overline}
\def\l{\left\langle}
\def\r{\right\rangle}
\def\as{\alpha_{s}}
\def\ycut{y_{\mbox{\tiny cut}}}
\def\yij{y_{ij}}
\def\epem{\ifmmode{e^+ e^-} \else{$e^+ e^-$} \fi}
\newcommand{\eeww}{$e^+e^-\rightarrow W^+ W^-$}
\newcommand{\qqQQ}{$q_1\bar q_2 Q_3\bar Q_4$}
\newcommand{\eeqqQQ}{$e^+e^-\rightarrow q_1\bar q_2 Q_3\bar Q_4$}
\newcommand{\eewwqqqq}{$e^+e^-\rightarrow W^+ W^-\ar q\bar q Q\bar Q$}
\newcommand{\eeqqgg}{$e^+e^-\rightarrow q\bar q gg$}
\newcommand{\eeqloop}{$e^+e^-\rightarrow q\bar q gg$ via loop of quarks}
\newcommand{\eeqqqq}{$e^+e^-\rightarrow q\bar q Q\bar Q$}
\newcommand{\eewwjjjj}{$e^+e^-\rightarrow W^+ W^-\rightarrow 4~{\rm{jet}}$}
\newcommand{\eeqqggjjjj}{$e^+e^-\rightarrow q\bar 
q gg\rightarrow 4~{\rm{jet}}$}
\newcommand{\eeqloopjjjj}{$e^+e^-\rightarrow q\bar 
q gg\rightarrow 4~{\rm{jet}}$ via loop of quarks}
\newcommand{\eeqqqqjjjj}{$e^+e^-\rightarrow q\bar q Q\bar Q\rightarrow
4~{\rm{jet}}$}
\newcommand{\eejjjj}{$e^+e^-\rightarrow 4~{\rm{jet}}$}
\newcommand{\jjjj}{$4~{\rm{jet}}$}
\newcommand{\qqbar}{$q\bar q$}
\newcommand{\ww}{$W^+W^-$}
\newcommand{\ar}{\rightarrow}
\newcommand{\sm}{${\cal {SM}}$}
\newcommand{\Dir}{\kern -6.4pt\Big{/}}
\newcommand{\Dirin}{\kern -10.4pt\Big{/}\kern 4.4pt}
\newcommand{\DDir}{\kern -8.0pt\Big{/}}
\newcommand{\DGir}{\kern -6.0pt\Big{/}}
\newcommand{\wwqqqq}{$W^+ W^-\ar q\bar q Q\bar Q$}
\newcommand{\qqgg}{$q\bar q gg$}
\newcommand{\qloop}{$q\bar q gg$ via loop of quarks}
\newcommand{\qqqq}{$q\bar q Q\bar Q$}

\def\st{\sigma_{\mbox{\scriptsize t}}}
\def\Ord{\buildrel{\scriptscriptstyle <}\over{\scriptscriptstyle\sim}}
\def\OOrd{\buildrel{\scriptscriptstyle >}\over{\scriptscriptstyle\sim}}
\def\jhep #1 #2 #3 {{JHEP} {\bf#1} (#2) #3}
\def\plb #1 #2 #3 {{Phys.~Lett.} {\bf B#1} (#2) #3}
\def\npb #1 #2 #3 {{Nucl.~Phys.} {\bf B#1} (#2) #3}
\def\epjc #1 #2 #3 {{Eur.~Phys.~J.} {\bf C#1} (#2) #3}
\def\zpc #1 #2 #3 {{Z.~Phys.} {\bf C#1} (#2) #3}
\def\jpg #1 #2 #3 {{J.~Phys.} {\bf G#1} (#2) #3}
\def\prd #1 #2 #3 {{Phys.~Rev.} {\bf D#1} (#2) #3}
\def\prep #1 #2 #3 {{Phys.~Rep.} {\bf#1} (#2) #3}
\def\prl #1 #2 #3 {{Phys.~Rev.~Lett.} {\bf#1} (#2) #3}
\def\mpl #1 #2 #3 {{Mod.~Phys.~Lett.} {\bf#1} (#2) #3}
\def\rmp #1 #2 #3 {{Rev. Mod. Phys.} {\bf#1} (#2) #3}
\def\cpc #1 #2 #3 {{Comp. Phys. Commun.} {\bf#1} (#2) #3}
\def\sjnp #1 #2 #3 {{Sov. J. Nucl. Phys.} {\bf#1} (#2) #3}
\def\xx #1 #2 #3 {{\bf#1}, (#2) #3}
\def\hepph #1 {{\tt hep-ph/#1}}
\def\preprint{{preprint}}

\def\beq{\begin{equation}}
\def\beeq{\begin{eqnarray}}
\def\eeq{\end{equation}}
\def\eeeq{\end{eqnarray}}
\def\a0{\bar\alpha_0}
\def\thrust{\mbox{T}}
\def\Thrust{\mathrm{\tiny T}}
\def\ae{\alpha_{\mbox{\scriptsize eff}}}
\def\ap{\bar\alpha_p}
\def\as{\alpha_{\mathrm{S}}}
\def\aem{\alpha_{\mathrm{EM}}}
\def\b0{\beta_0}
\def\cN{{\cal N}}
\def\cd{\chi^2/\mbox{d.o.f.}}
\def\Ecm{E_{\mbox{\scriptsize cm}}}
\def\ee{e^+e^-}
\def\enap{\mbox{e}}
\def\eps{\epsilon}
\def\ex{{\mbox{\scriptsize exp}}}
\def\GeV{\mbox{\rm{GeV}}}
\def\half{{\textstyle {1\over2}}}
\def\jet{{\mbox{\scriptsize jet}}}
\def\kij{k^2_{\bot ij}}
\def\kp{k_\perp}
\def\kps{k_\perp^2}
\def\kt{k_\bot}
\def\lms{\Lambda^{(n_{\rm f}=4)}_{\overline{\mathrm{MS}}}}
\def\mI{\mu_{\mathrm{I}}}
\def\mR{\mu_{\mathrm{R}}}
\def\MSbar{\overline{\mathrm{MS}}}
\def\mx{{\mbox{\scriptsize max}}}
\def\NP{{\mathrm{NP}}}
\def\pd{\partial}
\def\pt{{\mbox{\scriptsize pert}}}
\def\pw{{\mbox{\scriptsize pow}}}
\def\so{{\mbox{\scriptsize soft}}}
\def\st{\sigma_{\mbox{\scriptsize tot}}}
\def\ycut{y_{\mathrm{cut}}}
\def\slashchar#1{\setbox0=\hbox{$#1$}           
     \dimen0=\wd0                                 
     \setbox1=\hbox{/} \dimen1=\wd1               
     \ifdim\dimen0>\dimen1                        
        \rlap{\hbox to \dimen0{\hfil/\hfil}}      
        #1                                        
     \else                                        
        \rlap{\hbox to \dimen1{\hfil$#1$\hfil}}   
        /                                         
     \fi}                                         %
\def\etmiss{\slashchar{E}^T}
\def\Meff{M_{\rm eff}}
\def\Ord{\lsim}
\def\OOrd{\gsim}
\def\tq{\tilde q}
\def\tchi{\tilde\chi}
\def\lsp{\tilde\chi_1^0}

\def\gam{\gamma}
\def\ph{\gamma}
\def\be{\begin{equation}}
\def\ee{\end{equation}}
\def\bea{\begin{eqnarray}}
\def\eea{\end{eqnarray}}
\def\lsim{\:\raisebox{-0.5ex}{$\stackrel{\textstyle<}{\sim}$}\:}
\def\gsim{\:\raisebox{-0.5ex}{$\stackrel{\textstyle>}{\sim}$}\:}

\def\ino{\mathaccent"7E} \def\gluino{\ino{g}} \def\mgluino{m_{\gluino}}
\def\sqk{\ino{q}} \def\sup{\ino{u}} \def\sdn{\ino{d}}
\def\chargino{\ino{\omega}} \def\neutralino{\ino{\chi}}
\def\cab{\ensuremath{C_{\alpha\beta}}} \def\proj{\ensuremath{\mathcal P}}
\def\dab{\delta_{\alpha\beta}}
\def\zz{s-M_Z^2+iM_Z\Gamma_Z} \def\zw{s-M_W^2+iM_W\Gamma_W}
\def\prop{\ensuremath{\mathcal G}} \def\ckm{\ensuremath{V_{\rm CKM}^2}}
\def\aem{\alpha_{\rm EM}} \def\stw{s_{2W}} \def\sttw{s_{2W}^2}
\def\nc{N_C} \def\cf{C_F} \def\ca{C_A}
\def\qcd{\textsc{Qcd}} \def\susy{supersymmetric} \def\mssm{\textsc{Mssm}}
\def\slash{/\kern -5pt} \def\stick{\rule[-0.2cm]{0cm}{0.6cm}}
\def\h{\hspace*{-0.3cm}}

\def\ims #1 {\ensuremath{M^2_{[#1]}}}
\def\tw{\tilde \chi^\pm}
\def\tz{\tilde \chi^0}
\def\tf{\tilde f}
\def\tl{\tilde l}
\def\ppb{p\bar{p}}
\def\gl{\tilde{g}}
\def\sq{\tilde{q}}
\def\sqb{{\tilde{q}}^*}
\def\qb{\bar{q}}
\def\sqL{\tilde{q}_{_L}}
\def\sqR{\tilde{q}_{_R}}
\def\ms{m_{\tilde q}}
\def\mg{m_{\tilde g}}
\def\Gs{\Gamma_{\tilde q}}
\def\Gg{\Gamma_{\tilde g}}
\def\md{m_{-}}
\def\eps{\varepsilon}
\def\Ce{C_\eps}
\def\dnq{\frac{d^nq}{(2\pi)^n}}
\def\DR{$\overline{DR}$\,\,}
\def\MS{$\overline{MS}$\,\,}
\def\DRm{\overline{DR}}
\def\MSm{\overline{MS}}
\def\ghat{\hat{g}_s}
\def\shat{\hat{s}}
\def\sihat{\hat{\sigma}}
\def\Li{\text{Li}_2}
\def\bs{\beta_{\sq}}
\def\xs{x_{\sq}}
\def\xsa{x_{1\sq}}
\def\xsb{x_{2\sq}}
\def\bg{\beta_{\gl}}
\def\xg{x_{\gl}}
\def\xga{x_{1\gl}}
\def\xgb{x_{2\gl}}
\def\lsp{\tilde{\chi}_1^0}

\def\gluino{\mathaccent"7E g}
\def\mgluino{m_{\gluino}}
\def\squark{\mathaccent"7E q}
\def\msquark{m_{\mathaccent"7E q}}
\def\M{ \overline{|\mathcal{M}|^2} }
\def\utm{ut-M_a^2M_b^2}
\def\MiLR{M_{i_{L,R}}}
\def\MiRL{M_{i_{R,L}}}
\def\MjLR{M_{j_{L,R}}}
\def\MjRL{M_{j_{R,L}}}
\def\tiLR{t_{i_{L,R}}}
\def\tiRL{t_{i_{R,L}}}
\def\tjLR{t_{j_{L,R}}}
\def\tjRL{t_{j_{R,L}}}
\def\tg{t_{\gluino}}
\def\uiLR{u_{i_{L,R}}}
\def\uiRL{u_{i_{R,L}}}
\def\ujLR{u_{j_{L,R}}}
\def\ujRL{u_{j_{R,L}}}
\def\ug{u_{\gluino}}
\def\utot{u \leftrightarrow t}
\def\ar{\to}
\def\sqk{\mathaccent"7E q}
\def\sup{\mathaccent"7E u}
\def\sdn{\mathaccent"7E d}
\def\chargino{\mathaccent"7E \chi}
\def\neutralino{\mathaccent"7E \chi}
\def\slepton{\mathaccent"7E l}
\def\M{ \overline{|\mathcal{M}|^2} }
\def\cab{\ensuremath{C_{\alpha\beta}}}
\def\ckm{\ensuremath{V_{\rm CKM}^2}}
\def\zz{s-M_Z^2+iM_Z\Gamma_Z}
\def\zw{s-M_W^2+iM_W\Gamma_W}
\def\s22w{s_{2W}^2}

\newcommand{\cpmtwo}    {\mbox{$ {\chi}^{\pm}_{2}                    $}}
\newcommand{\cpmone}    {\mbox{$ {\chi}^{\pm}_{1}                    $}}

\begin{flushright}
{SHEP-08-31}\\
\today
\end{flushright}
\vskip0.1cm\noindent
\begin{center}
{\Large {\bf 
One-loop Electro-Weak Corrections to the \\[0.25cm]
 Event Orientation 
of $e^+e^-\to 3$~jet Processes\footnote{Work supported in 
part by the U.K. Science and Technology Facilities Council 
(STFC).}}}
\\[1.5cm]
{\large C.M. Carloni-Calame$^{1}$, S. Moretti$^{1}$, F. Piccinini$^{2}$ and D.A. Ross$^{1}$}\\[0.25 cm]
{\it $^1$ School of Physics and Astronomy, University of Southampton}\\
{\it Highfield, Southampton SO17 1BJ, UK}
\\[0.5cm]
{\em $^2$ INFN - Sezione di Pavia,
    Via Bassi 6, 27100 Pavia,
Italy}
\\[0.5cm]
\end{center}

\begin{abstract}
{\small
\noindent
We compute the full one-loop Electro-Weak (EW)
contributions of ${\cal O}(\alpha_{\rm S}\alpha_{\rm{EM}}^3)$ 
entering the electron-positron into a quark-antiquark pair plus one gluon cross section 
at the $Z$ peak and LC energies in presence of polarisation of the initial 
state
and by retaining the event orientation of the final state.
We include both factorisable and non-factorisable virtual corrections,
photon bremsstrahlung but not the real emission of $W^\pm$
and $Z$ bosons. Their importance for the final state orientation 
is illustrated for beam polarisation setups achieved at SLC and foreseen at 
ILC and CLIC.}
\end{abstract}

\section{Introduction}
\label{Sec:Intro}
There are innumerable tests of Quantum Chromo-Dynamics (QCD) that the $e^+e^-\to q\bar qg$ cross section can enable: it provides direct 
evidence for the existence of the gluon (the QCD gauge boson), it also allows one to measure its spin and to confirm its non-abelian nature,
it permits the measurement of  the QCD coupling constant ($\alpha_{\rm S}$), it is sensitive to the Casimir form factors of QCD. Furthermore, if the
flavour of the final state quark can be tagged (as  can efficiently be done for the case of $b$-quarks, thanks to $\mu$-vertex or
high-$p_T$ lepton 
techniques), one can use $e^+e^-\to b\bar bg$ samples to verify the flavour independence of $\alpha_{\rm S}$ and to measure the 
$b$-quark (running) mass. 

Thorough tests of QCD have  been performed over the years and at many colliders (PETRA, TRISTAN, SLC/LEP and
LEP2 \cite{QCD}) through  $e^+e^-\to q\bar qg$ and many more are foreseen at future ones, such as the ILC \cite{Bethke:1993ym} 
or CLIC \cite{CLIC}.
In fact, old $e^+e^-\to q\bar qg$ data are currently been revisited \cite{JADE} in the light of the recently available 
Next-to-Next-to-Leading-Order NNLO QCD corrections \cite{QCD2Loops}, which can now supplement the 
 NLO results of \cite{ERT}
(see also Refs.~\cite{EVENT}--\cite{EVENT2} for their implementation in numerical programs).
However, as already emphasised in Refs.~\cite{oldpapers,poleeEW}, if ${\cal O}(\alpha_{\rm S}^3\alpha_{\rm{EM}}^2)$ results (NNLO QCD)
are of experimental relevance, so are those of ${\cal O}(\alpha_{\rm S}\alpha_{\rm{EM}}^3)$  (NLO EW),
 since   at the energies of the aforementioned
colliders, ${\cal O}(\alpha_{\rm S}^2)\approx {\cal O}(\alpha_{\rm{EM}})$.   

If full event orientation is retained in $e^+e^-\to q\bar qg$, one can further study certain polar  and azimuthal angle asymmetries, which
are strongly sensitive to parity-violating effects and thus represent a new search-ground for anomalous contributions that
can be explored experimentally \cite{BurrowsOsland}--\cite{Laermann}. Such asymmetries are observed 
if beam polarisation can be exploited, as
 was the case at SLC and will certainly be possible at the ILC and CLIC. Finally, in the presence of polarised electrons and/or positrons, 
left-right asymmetries
can also be studied. In these respects, while NLO and NNLO QCD effects can be the source of non-trivial asymmetry effects,
one expects genuine parity-violating ones to  occur naturally in NLO EW corrections. 
It is the purpose of this note to show that this is indeed the case, and at a sizable level
 given current experimental uncertanties,
in both the unpolarised  and polarised $e^+e^-\to q\bar q g$ cross section.

The paper is organised as follows. Section \ref{Sec:Calculation} briefly 
describes our calculation.
Section \ref{Sec:Results} presents our numerical results whilst Section \ref{Sec:Summary}
 summarises our main conclusions.

\section{Calculation}
\label{Sec:Calculation}

The procedures adopted to carry out our computation have been described in the first paper of \cite{poleeEW},
to which we refer the reader for the most technical details. Here, we would only like to point out that,
in anticipation of an electron-positron collider in which either or both the incoming beams 
can be polarised, we have inserted a helicity projection operator 
into the electron line and obtained separate results for left-handed
($e_L$) and right-handed ($e_R$) incoming
electrons\footnote{As we are taking massless incoming fermions,
 the helicity of the positron
is simply the opposite to that of the electron.}. As already intimated in the Introduction,
for genuinely weak interaction terms, this is of particular interest, since such corrections violate 
parity conservation. Unfortunately, this occurs already at tree-level \cite{BurrowsOsland}, owing to the contribution from 
exchange of a $Z$ boson, but these higher order 
corrections are also peculiarly dependent on the incoming lepton helicity and thus one would expect the two parity-violating effects be 
distinguishable after the collection of sufficient events. Also notice that,
as well as  the electron/positron mass, we also have neglected the masses of the quarks throughout. However,  
whenever there is a $W^\pm$ boson in the virtual loops, account has to be taken of the mass of the virtual 
top (anti)quark, which we have done. 

Before proceeding to show our results, we should mention the numerical
parameters used for our simulations. We have
taken the top (anti)quark to have a
mass $m_t = 171.6$ GeV. The $Z$ mass used was $M_Z = 91.18$ GeV and was
related to the $W^\pm$ mass, $M_W$, via the Standard Model (SM) formula
$M_W = M_Z \cos \theta_W$, where $\sin^2 \theta_W =$ 0.222478. The $Z$
width was $\Gamma_Z = 2.5$ GeV.
Also notice that, where relevant, Higgs contributions were included
with $M_H=115$ GeV. For the strong
coupling constant, $\alpha_{\rm{S}}$, we have used the two-loop expression with
$\Lambda^{(n_f=4)}_{\rm{QCD}}=0.325$ GeV in the
$\overline{\mathrm{MS}}$ scheme, yielding
$\alpha_{\rm{S}}^{\overline{\mathrm{MS}}}(M_Z^2)=0.118$.

As for the jet definition,  partonic momenta are clustered into jets 
according to the
Cambridge jet algorithm~\cite{cambridge} (e.g., when $y_{ij}<y_{cut}$
with $y_{cut}=0.001$), the jets are required to lie in the central
detector region $30^\circ<\theta_{\mathrm{jets}}<150^\circ$ and we require that
the invariant mass of the jet system $M_{q\bar qg}$ is larger than
$0.75\times\sqrt{s}$.
If a real photon is present in the final state, it is
clustered according to the same algorithm, but we require that
at least three ``hadronic'' jets are left at the end
(i.e., events in which the photon is resolved are rejected)\footnote{As explained in the first paper in \cite{poleeEW}, 
this serves a twofold purpose. On the
one hand, from the experimental viewpoint, a resolved (energetic and isolated)
single photon is never treated as a jet. On the other hand, from a theoretical
viewpoint, this enables us to remove divergent contributions appearing 
whenever an unresolved (soft and/or collinear) gluon is emitted, since
we are not computing here the one-loop
 ${\cal O}(\alpha_{\rm S}\alpha_{\rm{EM}}^3)$
QCD corrections to $e^+e^-\to q\bar q\gamma$.}. 

We will look at the cases of: (i) fully inclusive cross section, where no
parton state can be identified from the jets; 
(ii) semi-inclusive cross section, where,
e.g., the quark is assumed to be tagged and the gluon is taken to be
the least energetic jet in the event; (iii) (fully) exclusive cross section,
where each parton can be identified with a jet. Recall that, al least
in the case of $b$-quarks, both the flavour (e.g., via a $\mu$-vertex device) 
and charge  (e.g., 
via the emerging lepton charge or the jet-charge method) of 
a quark can be extracted from a jet.
 (In all such cases, for sake of
illustration, we take the efficiency to be unity.) 

In order to show the behaviour of the EW corrections we are calculating, other than
scanning in the collider energy, we have considered here the three discrete
values of $\sqrt{s}=M_Z$ (for LEP/SLC, but also in view of a GigaZ option of a future LC), 
$\sqrt{s}=350$ (the $t\bar t$ threshold of a ILC and/or CLIC) GeV and $\sqrt{s}=1$ TeV  (as representative of
the so-called `Sudakov regime' \cite{Ciafaloni:1999xg} which can be realised at both ILC and CLIC). For reference, concerning beam polarisation, we will
adopt 70\% electron polarisation at $\sqrt{s}=M_Z$ (thus emulating the SLC
configuration) and 100\% at $\sqrt{s}=350$ GeV and 1 TeV (thus emulating a
possible ILC/CLIC configuration)\footnote{Recall that, in our case, since
we take all external fermions to be massless, the positron has
opposite helicity to that of the electron.}.

\section{Numerical Results}
\label{Sec:Results}

\begin{figure}\begin{center}{
\includegraphics[width=12.5cm]{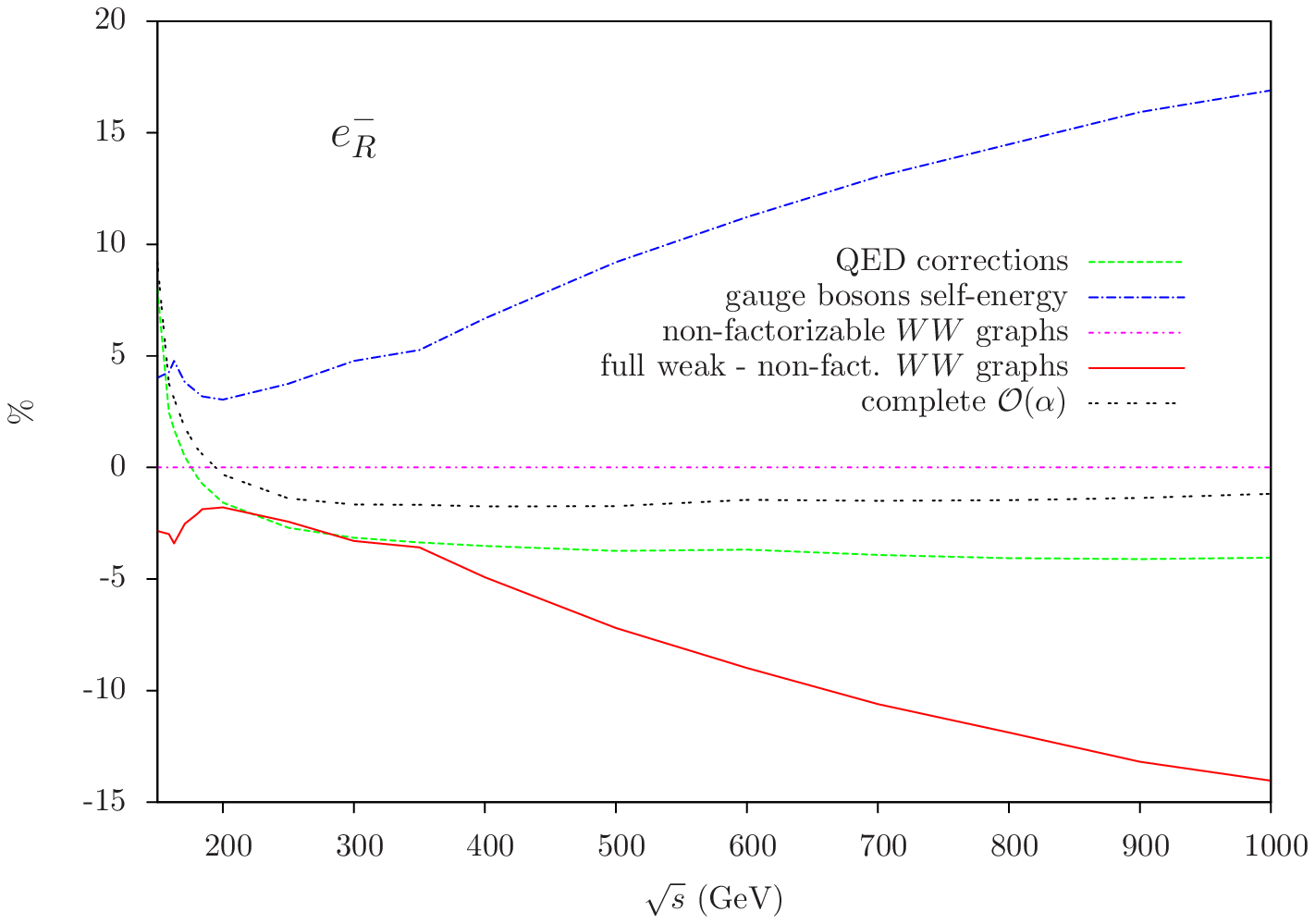}
\includegraphics[width=12.5cm]{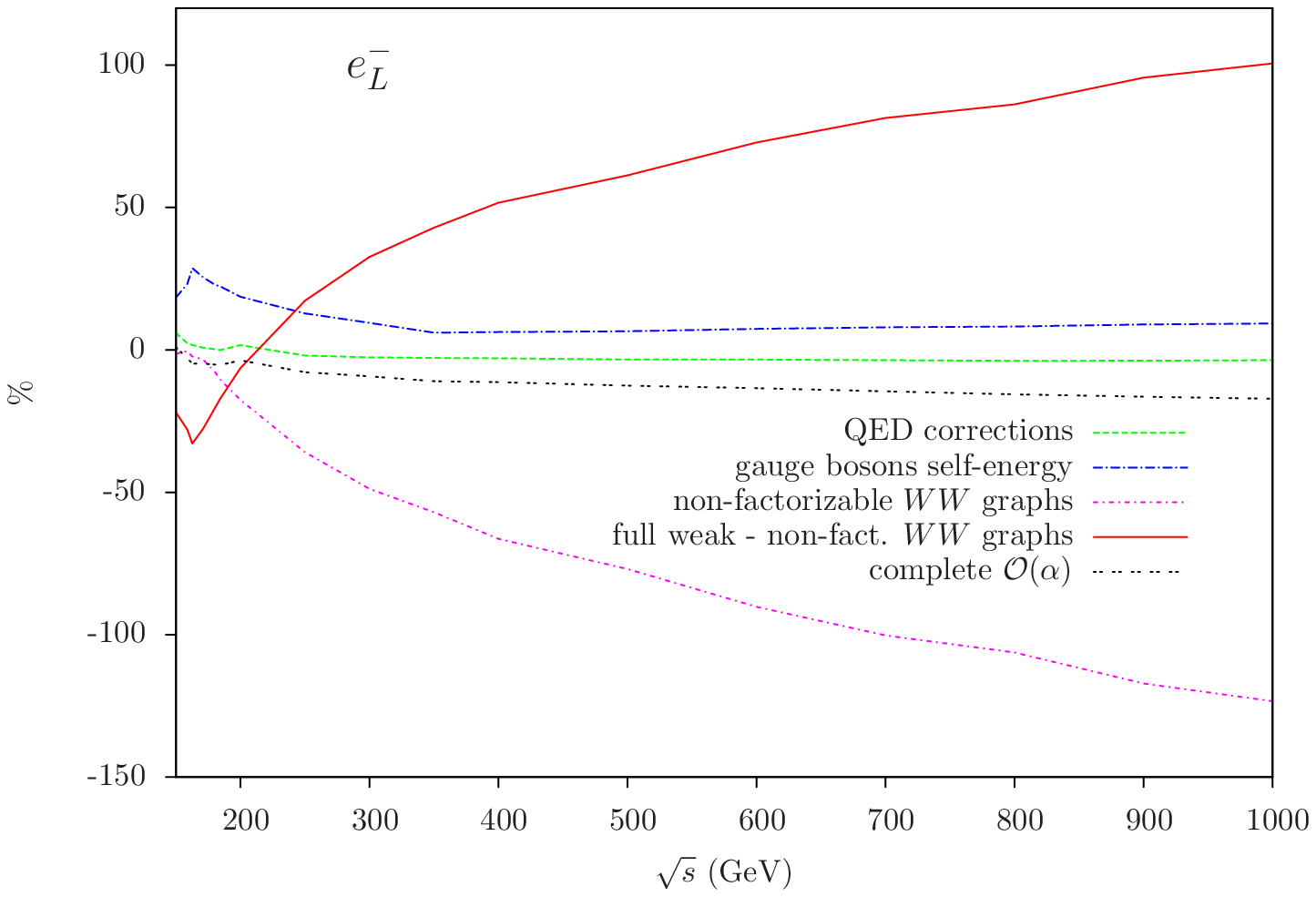}
\caption{Top/Bottom is for $e^-_R/e^-_L$: 
Relative effect on the integrated cross section due to
  different contributions to the order $\alpha\equiv
\alpha_{\rm{EM}}$ correction, as a
  function of the CM energy.}
\label{scan}}\end{center}\end{figure}

Before proceeding to investigate the spatial orientation of the $e^+e^-\to
q\bar qg$ differential cross section through the 
${\cal O}(\alpha_{\rm S}\alpha_{\rm{EM}}^3)$ in presence of
one-loop EW corrections of ${\cal O}(\alpha_{\rm{EM}})$, it is instructive
to see the effects of the various components of these corrections
 on the  tree-level ${\cal O}(\alpha_{\rm S}\alpha_{\rm{EM}}^2)$  result
for the integrated cross section, as a function of the collider energy.
This is done in Fig.~\ref{scan}, for the two electron polarisations
separately.  The curves represent the effects of
the QED (virtual and real) corrections only, the gauge
bosons self-energy corrections, the non-factorisable graphs involving
four- and five-point functions with $WW$
exchange\footnote{This is a gauge invariant subset of the complete
corrections.},
the weak corrections with the non-factorisable $WW$ graphs removed
 and the sum of the previous
ones. Notice that
the total effect depends strongly on the electron polarisation
state. If the electron is right-handed, there are no
contributions from the non-factorisable $WW$ graphs. Furthermore,
there are strong cancellations between the gauge
bosons self-energy terms (which are increasingly positive) and the full weak
corrections with the  non-factorisable $WW$ graphs omitted,
(which are increasingly negative). Here,
the QED terms are never very large. Overall, the full  
${\cal O}(\alpha_{\rm{EM}})$
correction is small, as it is always below the $-2\%$ level.
In contrast, overall effects are very large in the case of left-handed
electrons, also displaying the typical Sudakov enhancement at
large energies: e.g., at 1 TeV, the full ${\cal O}(\alpha_{\rm{EM}})$
correction can reach $-20\%$. Here, the dominant terms, again of opposite
signs, are the full weak
minus non-factorisable $WW$ component (increasingly positive) 
and the non-factorisable $WW$ terms (increasingly negative),
with the QED effects being at the (positive) percent level
and the gauge boson self-energies yielding up to $+25\%$ corrections.

\begin{figure}\begin{center}{
\includegraphics[width=12.5cm]{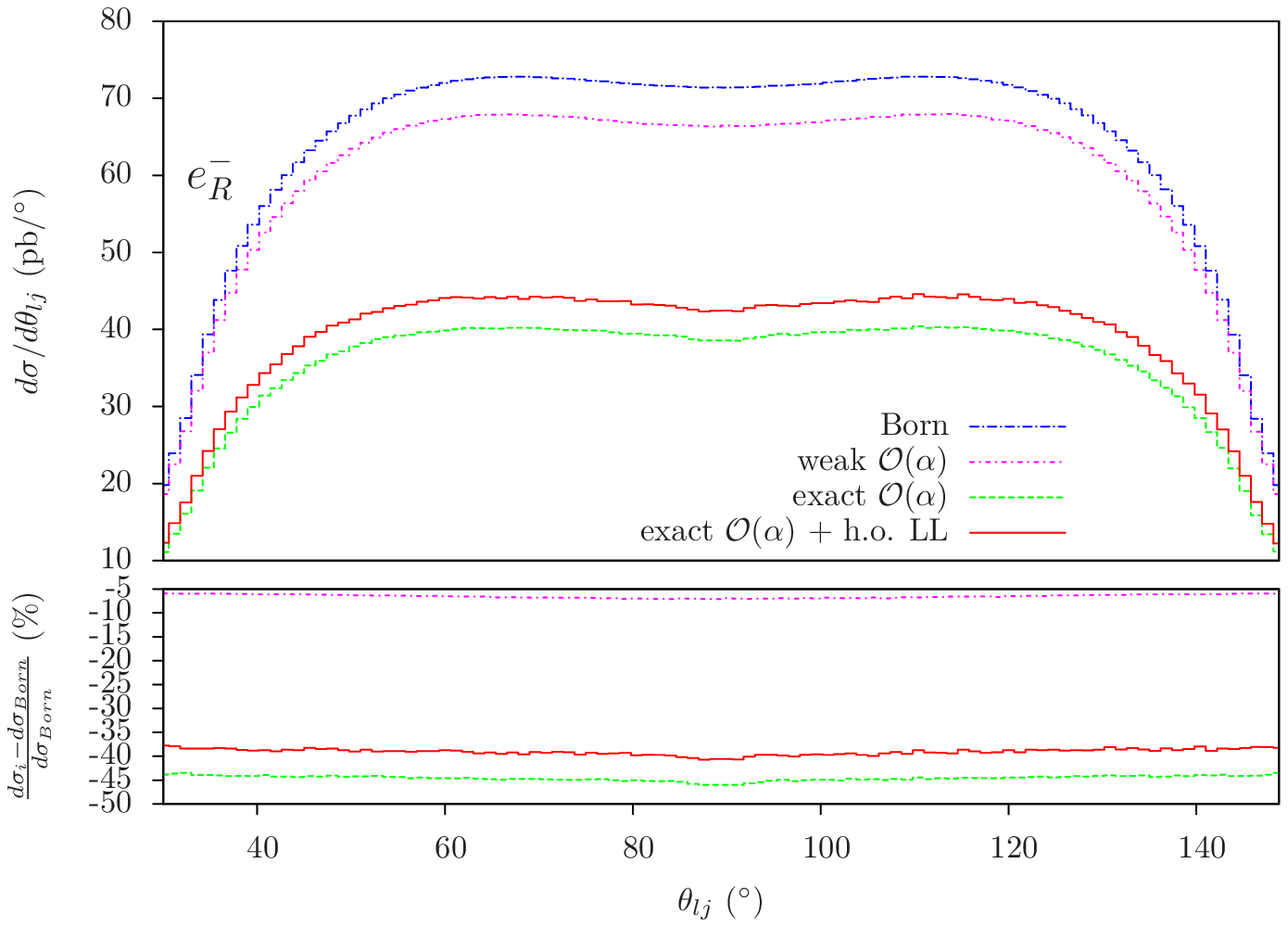}
\includegraphics[width=12.5cm]{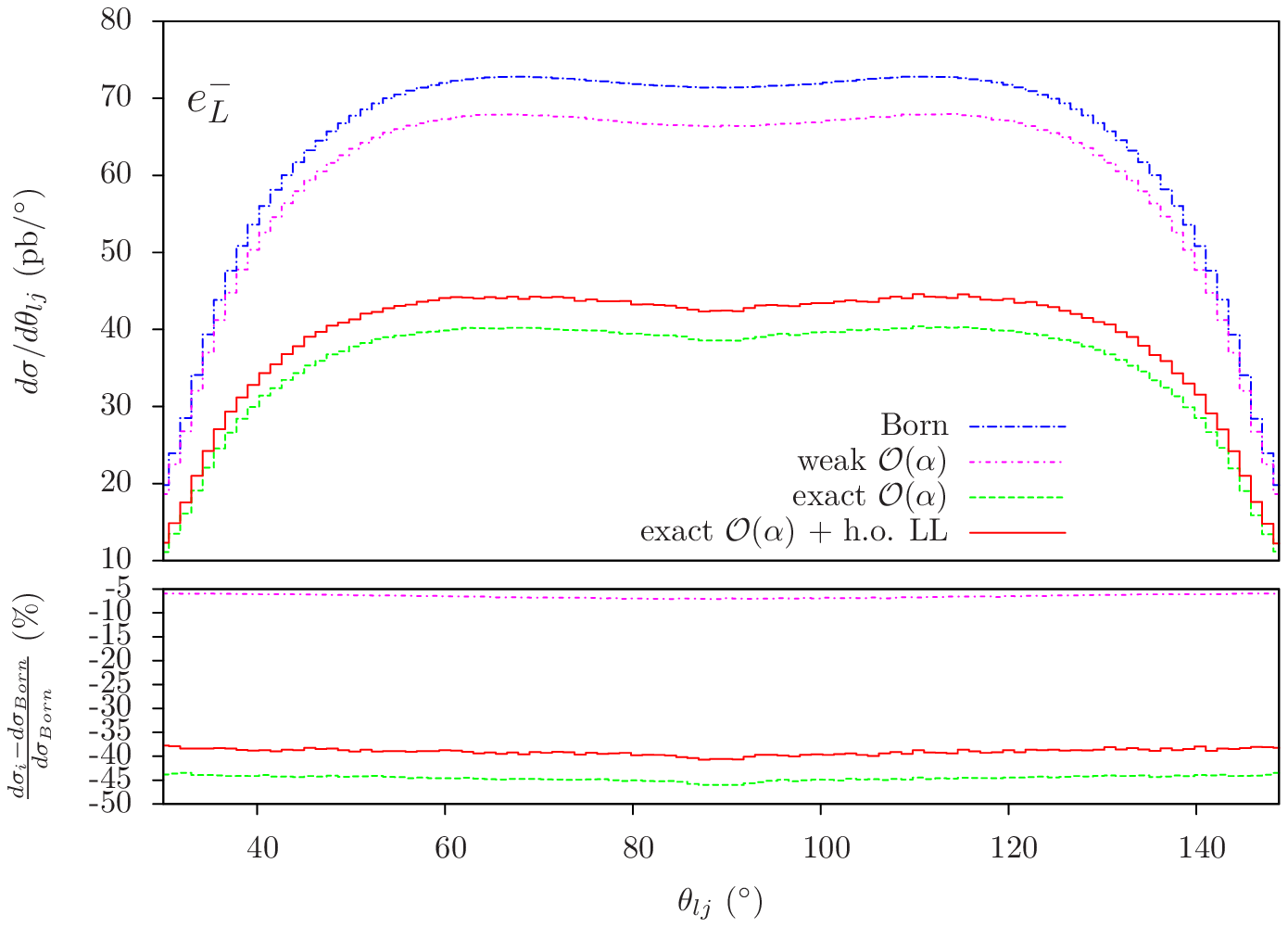}
\caption{Top/Bottom is for $e^-_R/e^-_L$: $\frac{d\sigma}{d\theta_{lj}}$ distribution at the $Z$ peak. (See the main text for the definition of this variable.)}
\label{ajl-peak}}\end{center}\end{figure}

\begin{figure}\begin{center}{
\includegraphics[width=12.5cm]{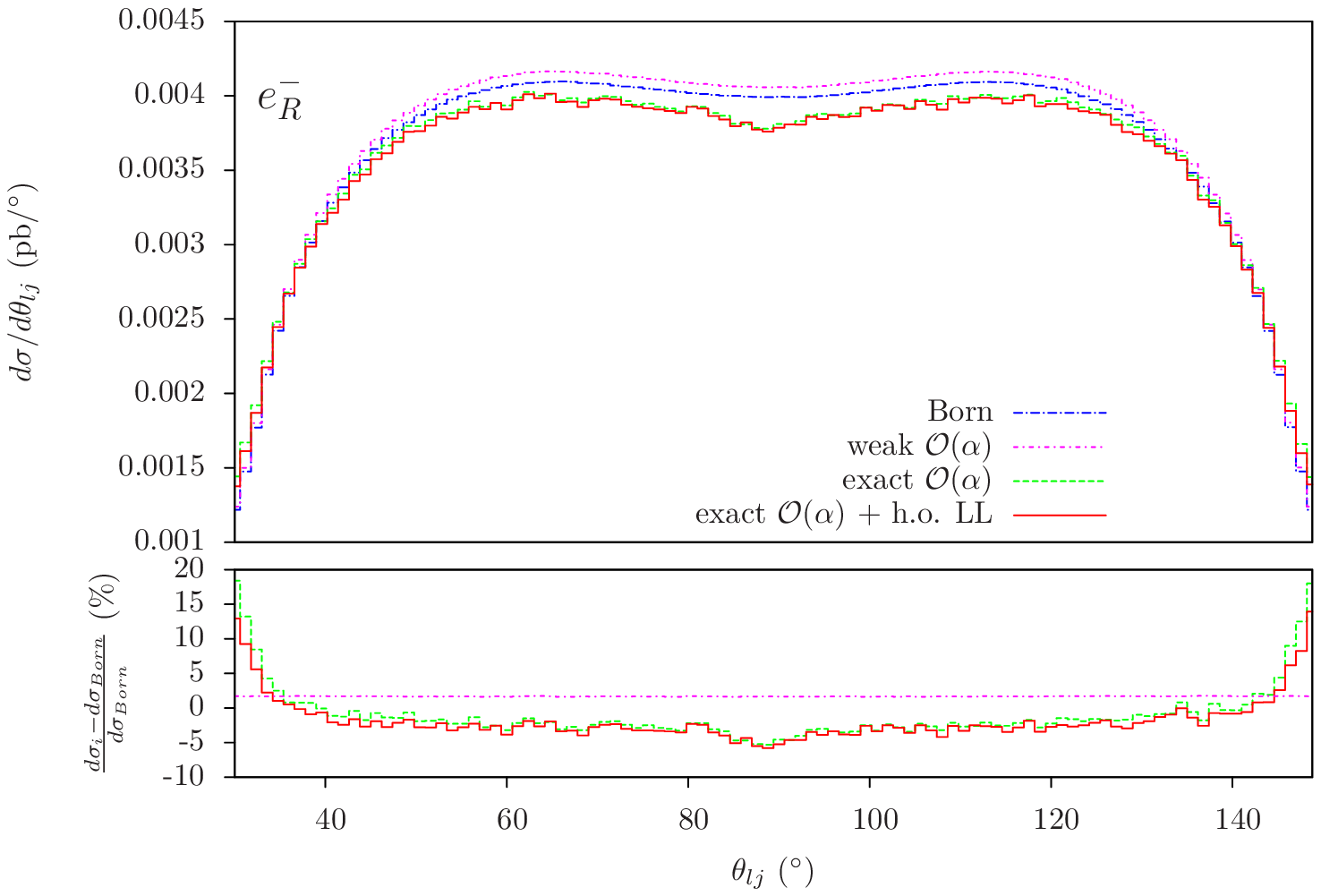}
\includegraphics[width=12.5cm]{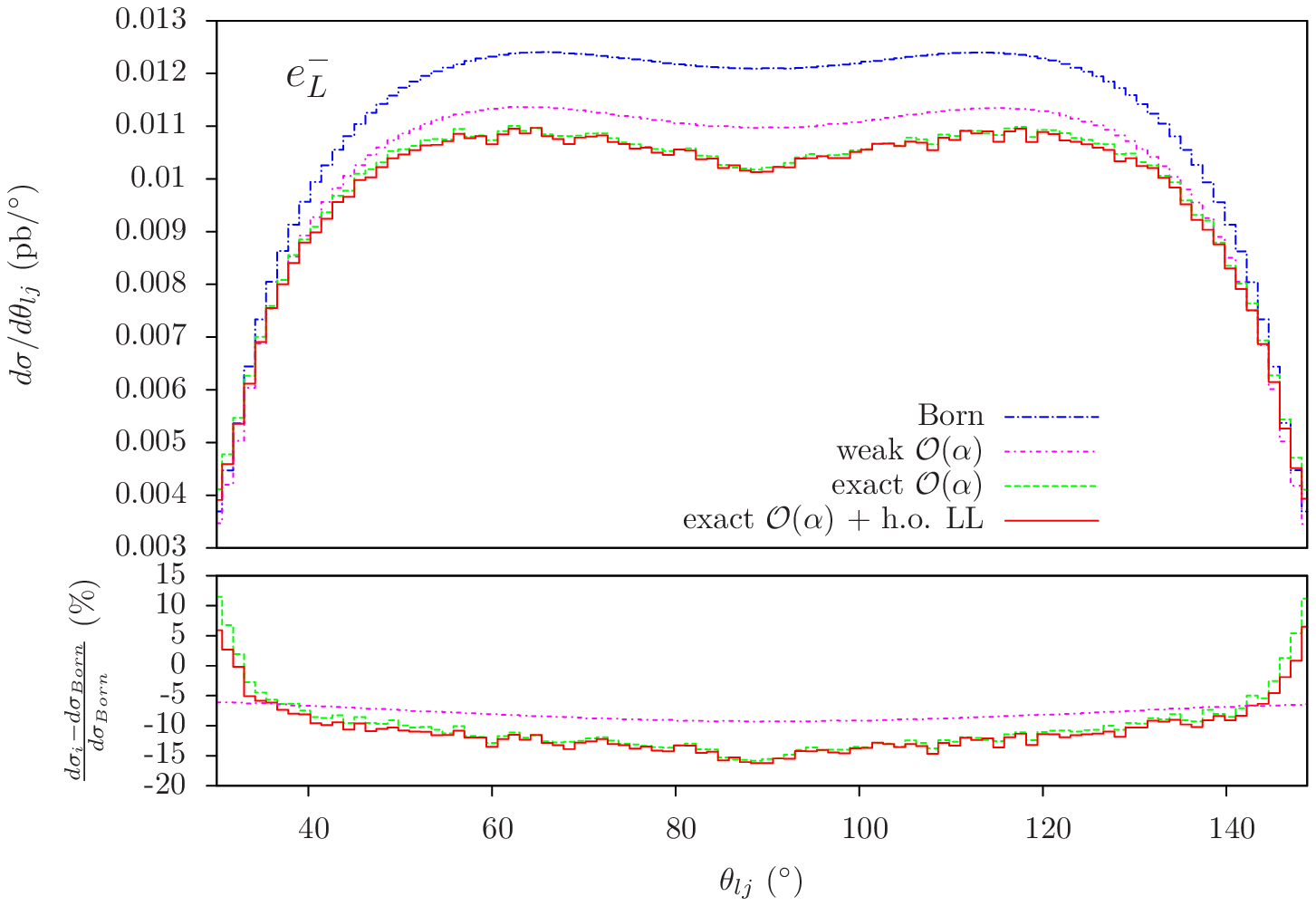}
\caption{Top/Bottom is for $e^-_R/e^-_L$: $\frac{d\sigma}{d\theta_{lj}}$ distribution at 350 GeV.  (See the main text for the definition of this variable.)}
\label{ajl-350}}\end{center}\end{figure}

\begin{figure}\begin{center}{
\includegraphics[width=12.5cm]{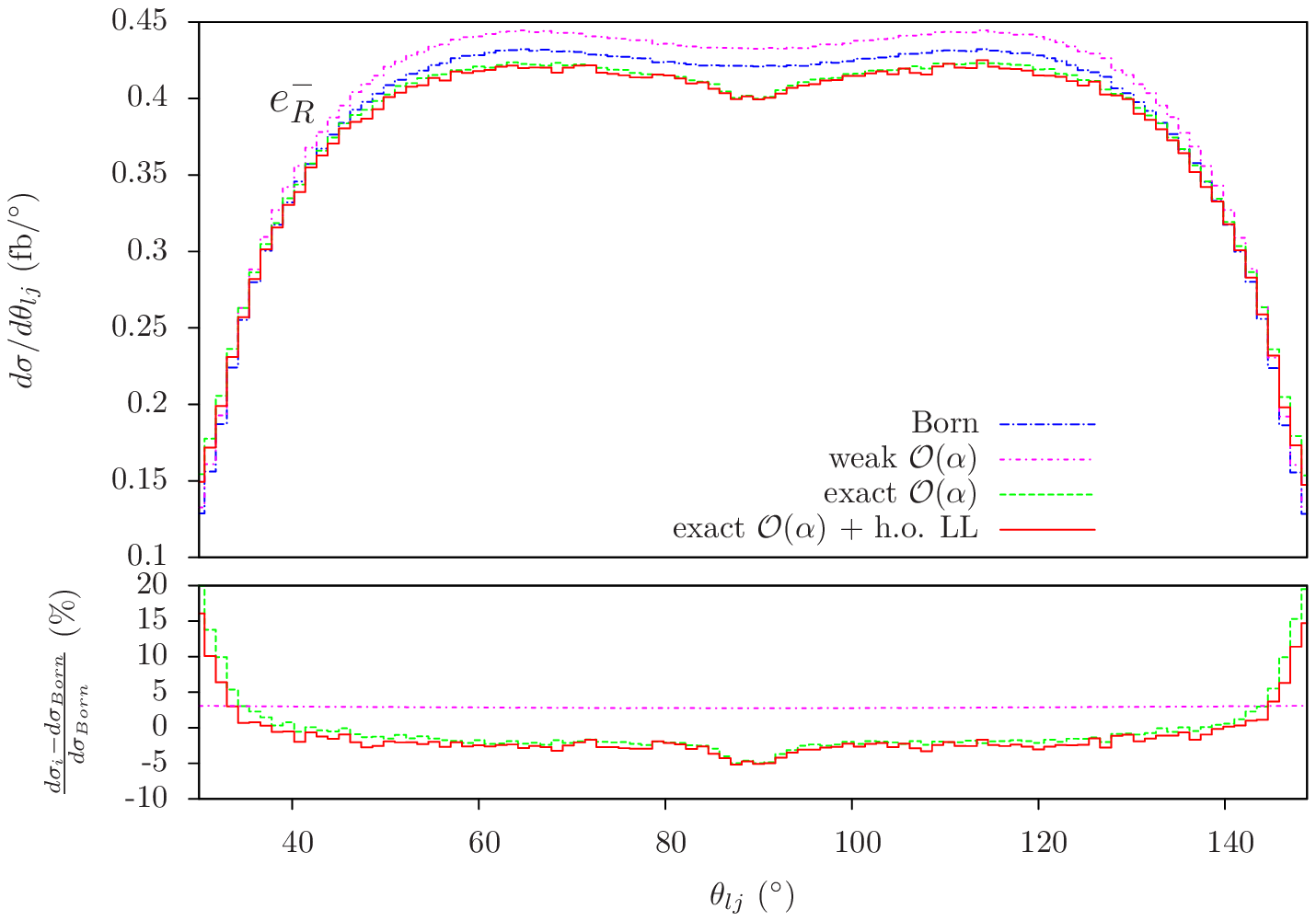}
\includegraphics[width=12.5cm]{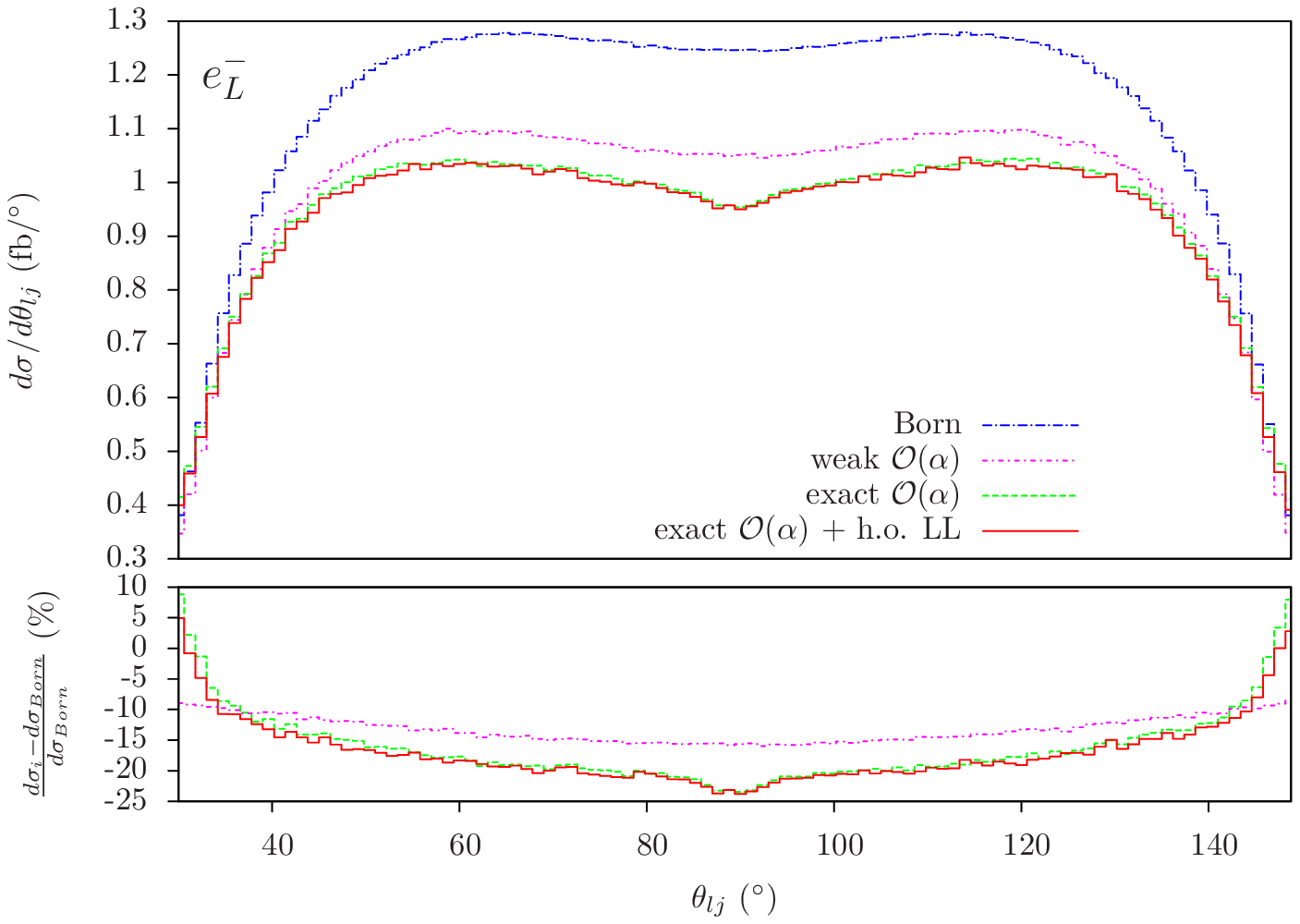}
\caption{Top/Bottom is for $e^-_R/e^-_L$: $\frac{d\sigma}{d\theta_{lj}}$ distribution at 1 TeV.  (See the main text for the definition of this variable.)}
\label{ajl-1000}}\end{center}\end{figure}

We start our differential analysis by assuming a fully inclusive 
setup, whereby
jets are labelled only in terms of their energies,
$E_3\le E_2\le E_1$, and the event orientation is
 identified uniquely by adopting as polar and azimuthal
variables
(see Ref.~\cite{BurrowsOsland}) those defined
as the   
angle of the fastest jet with respect to the electron beam direction
(denoted by $\theta_{ij}$)
and the one whose cosine satisfies (with obvious meaning of the 
vector symbols)
\begin{equation}
\cos\chi'' =  {{\vec 1}\times {\vec 3} \over{
| {\vec 1}\times {\vec 3} |}}
\cdot {{\vec 1} \times \vec{e^-} \over {
| {\vec 1} \times {\vec{e^-}}|}},
\label{chi2nd}
\end{equation}
respectively. At the $Z$ pole, EW corrections of ${\cal O}(\alpha_{\rm{EM}})$
to the polar angle are substantial, up
to $-40\%$ at fixed order and dominated by the QED component, as the purely 
weak part, involving only the exchange of $W^\pm$ and/or $Z$ bosons, accounts
only for a $-5\%$ or so. The QED effects are tamed by the inclusion 
of even higher order Leading Logarithmic (LL) effects (see \cite{poleeEW} for
the details of the implementation),
though they decrease only slightly, to $-35\%$. Such large effects are
mainly driven by the Initial State Radiation (ISR) affecting the line-shape
of the cross section around the resonance and are
essentially independent of the polarisation state of the electron. With
increasing collider energy, for the case of left-handed electrons, 
the overall corrections diminish
in absolute size, to no more than $-25\%$ (at $\sqrt s=1$ TeV), even becoming
positive in the forward/backward directions. Such
a decrease affects the QED component. At the same time, owing
to the onset of the Sudakov regime, the effects due to the weak component
of the ${\cal O}(\alpha_{\rm{EM}})$ corrections increase in size, to $-15\%$ 
or so at 1 TeV (in the central region). In contrast with the case of
low energy, for high values of $\sqrt{s}$, the  ${\cal O}(\alpha_{\rm{EM}})$
effects induced by right-handed electrons are different, being smaller
by about  factor of 3 when the corrections are negative and
larger by about a factor of 3 when the corrections are positive,
for both $\sqrt s=350$ GeV and  $\sqrt s = 1 $ TeV.
 For the
${\cal O}(\alpha_{\rm{EM}})$ purely weak part, for right-handed electrons,
it can be seen that
corrections become positive and somewhat smaller, 
in this case independent of whether the collider energy is at the
$t\bar t$ threshold or in the TeV regime. We display all these patterns in 
Figs.~\ref{ajl-peak}--\ref{ajl-1000}.

\begin{figure}\begin{center}{
\includegraphics[width=12.5cm]{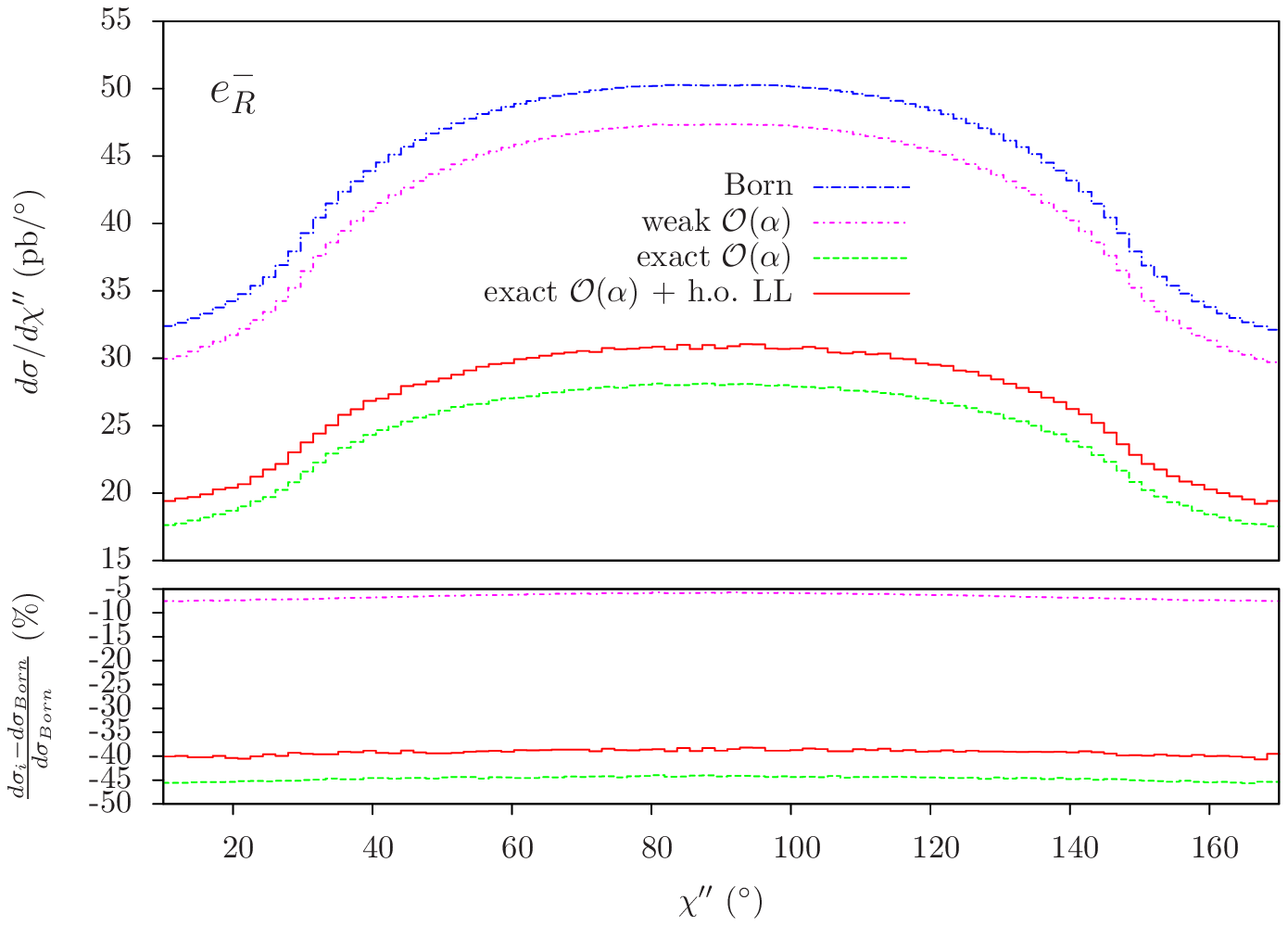}
\includegraphics[width=12.5cm]{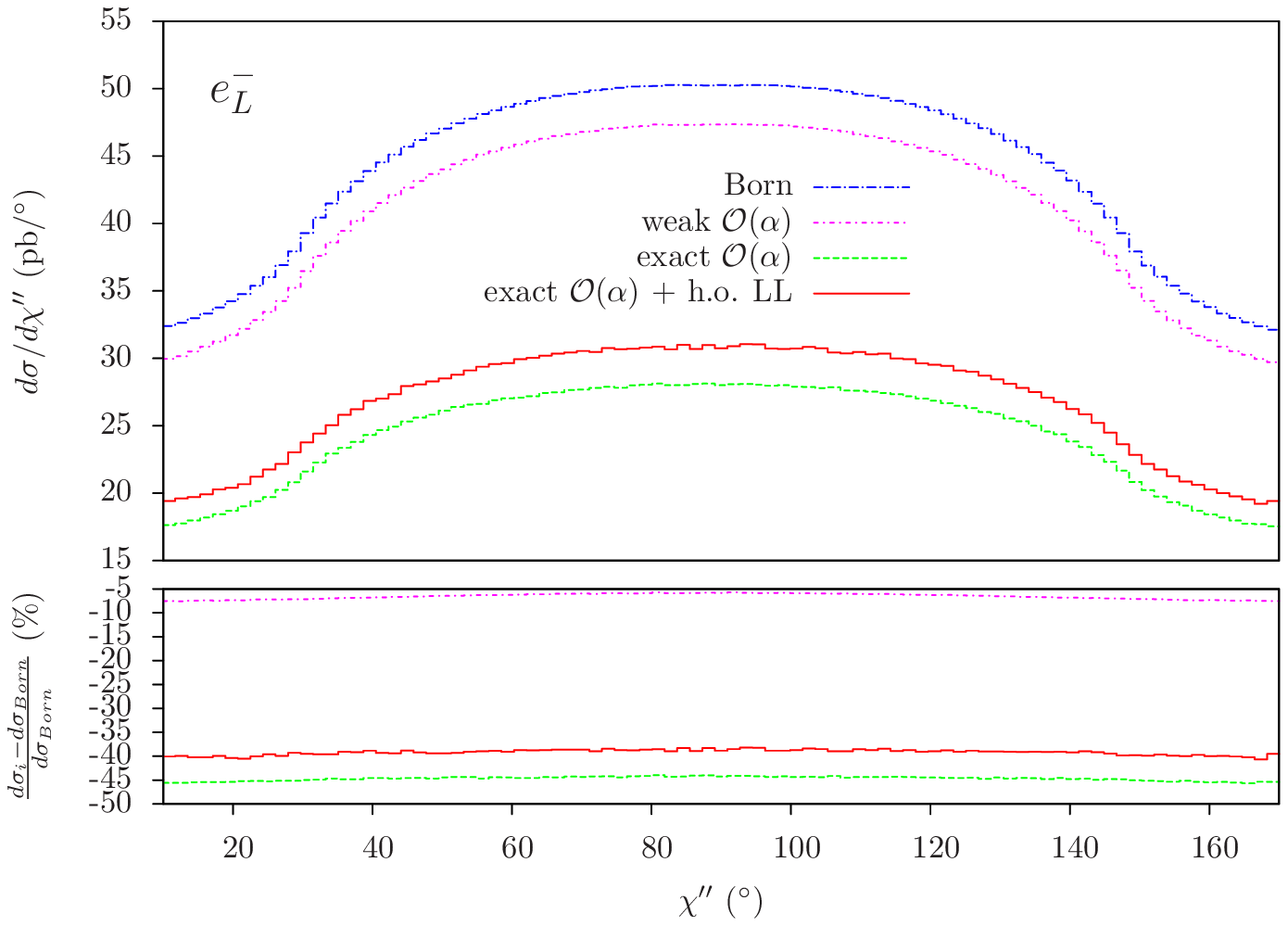}
\caption{Top/Bottom is for $e^-_R/e^-_L$: $\frac{d\sigma}{d\chi''}$ distribution at the $Z$ peak.  (See the main text for the definition of this variable.)}
\label{chi2nd-peak}}\end{center}\end{figure}

\begin{figure}\begin{center}{
\includegraphics[width=12.5cm]{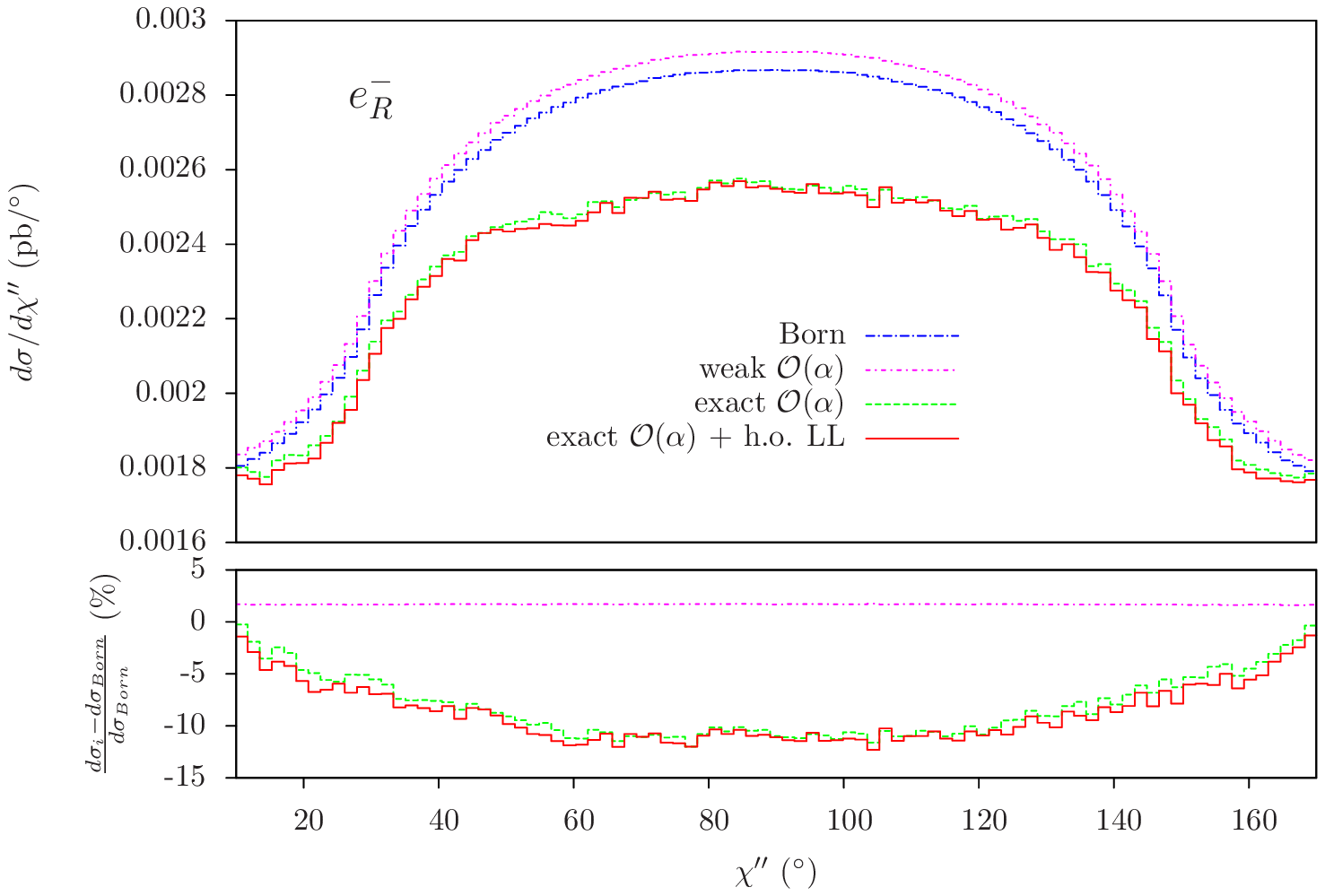}
\includegraphics[width=12.5cm]{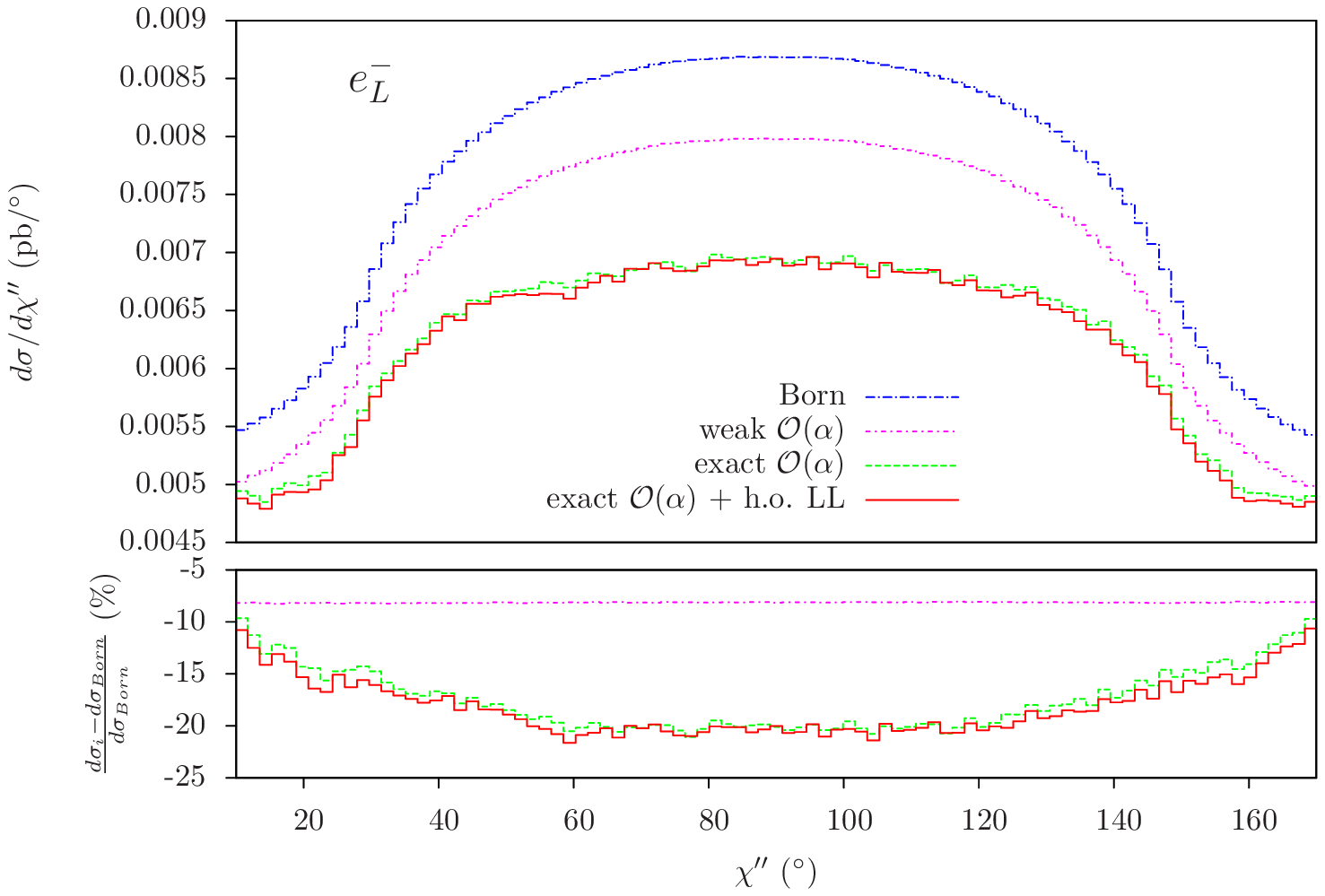}
\caption{Top/Bottom is for $e^-_R/e^-_L$: $\frac{d\sigma}{d\chi''}$ distribution at 350 GeV.  (See the main text for the definition of this variable.)}
\label{chi2nd-350}}\end{center}\end{figure}

\begin{figure}\begin{center}{
\includegraphics[width=12.5cm]{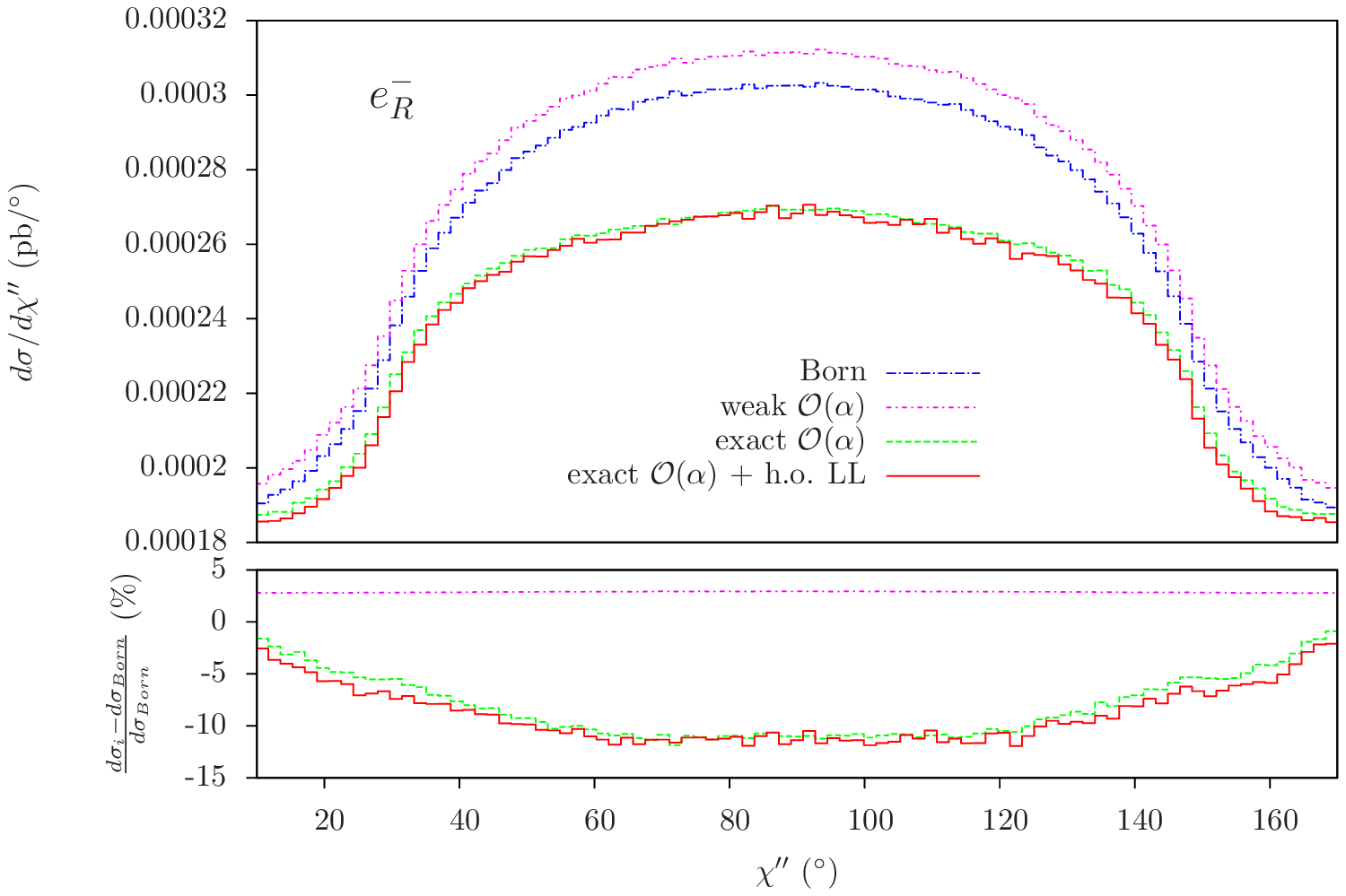}
\includegraphics[width=12.5cm]{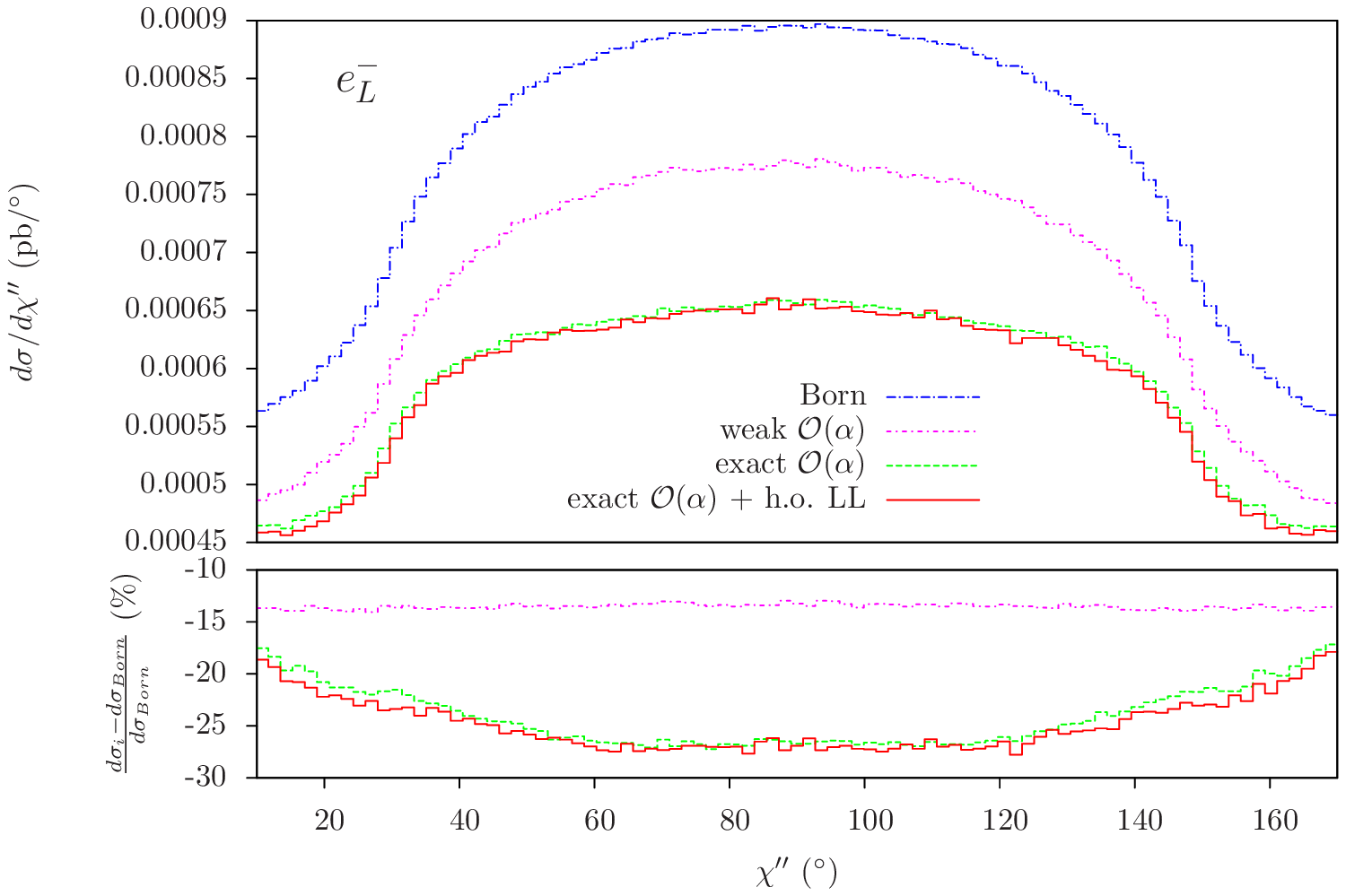}
\caption{Top/Bottom is for $e^-_R/e^-_L$: $\frac{d\sigma}{d\chi''}$ distribution at 1 TeV.  (See the main text for the definition of this variable.)}
\label{chi2nd-1000}}\end{center}\end{figure}

Unlike the case of the polar angle, the corrections to the differential
cross sections with resepct to  the azimuthal angle
at the $Z$ peak are essentially flat. Their hierarchy and size, however,
are the same for both variables. In fact, at fixed ${\cal O}(\alpha_{\rm{EM}})$,  
QED corrections are always the leading ones whilst the weak effects are subleading,
the former being a factor of 9 or so larger than the latter, both being
negative and summing up to $-45\%$. LL resummation effects in QED reduce the overall
corrections to $-40\%$. Again, at this low energy, there is no dependence
of the ${\cal O}(\alpha_{\rm{EM}})$ 
 EW effects on the electron polarisation. As $\sqrt s$ increases,
QED effects are no longer constant, as they are maximal (and negative)
for $\chi''=90^{\rm o}$, about $-15\%(-5\%)$ at 350 GeV and $-20\%(-5\%)$ 
at 1 TeV, for left-handed(right-handed) electrons, with minimal effects
due to the QED LL resummation. The purely weak corrections are
instead constant throughout the entire angular range, approximately 
$-6\%(-8\%)[-12\%]$ at $\sqrt s=M_Z(350~{\rm GeV})[1~{\rm TeV}]$, for
left-handed electrons. For right-handed ones, the corresponding numbers
are $-9\%(+2\%)[+3\%]$ or so. All this is shown in 
Figs.~\ref{chi2nd-peak}--\ref{chi2nd-1000}.

\begin{figure}\begin{center}{
\includegraphics[width=12.5cm]{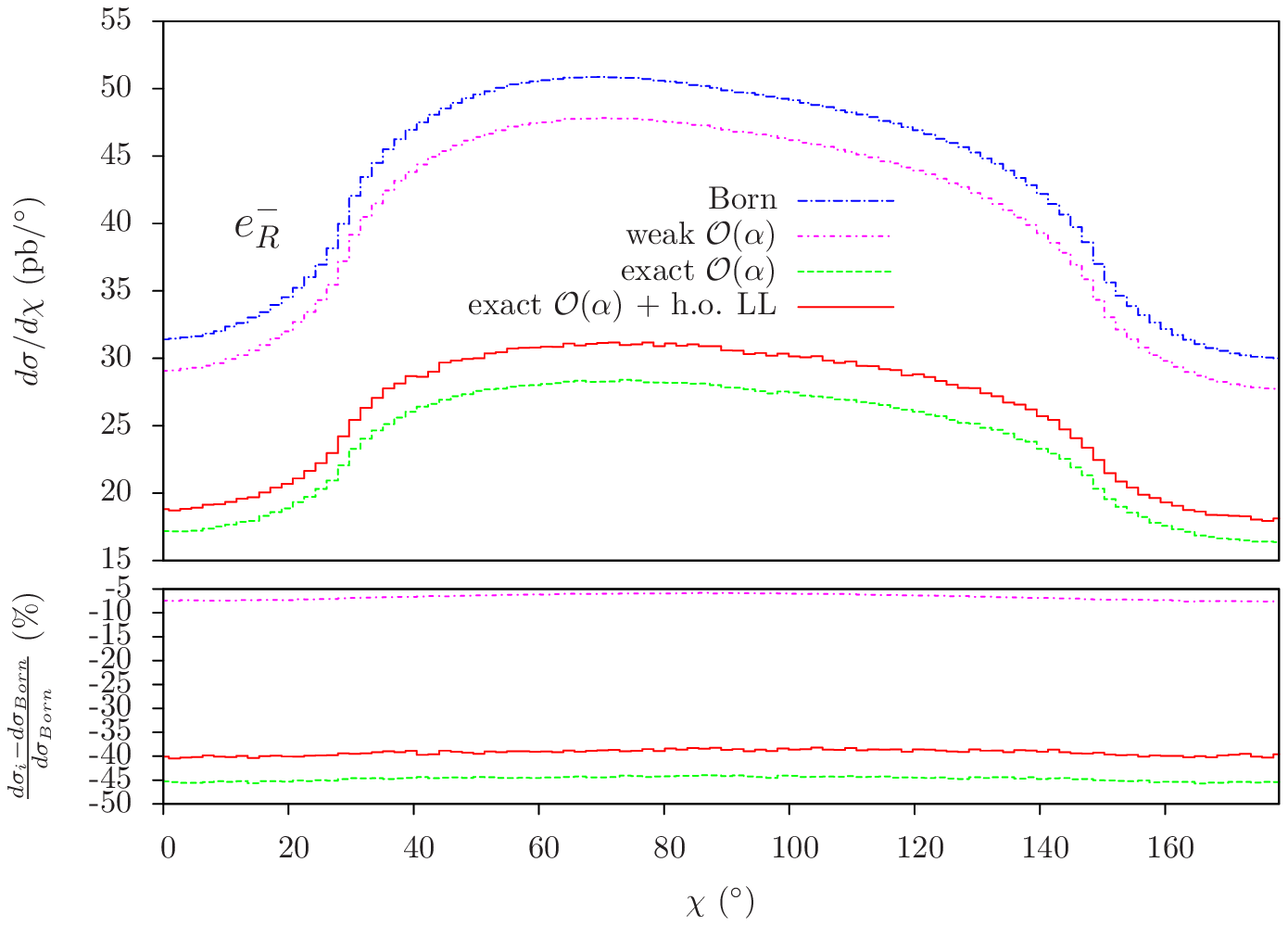}
\includegraphics[width=12.5cm]{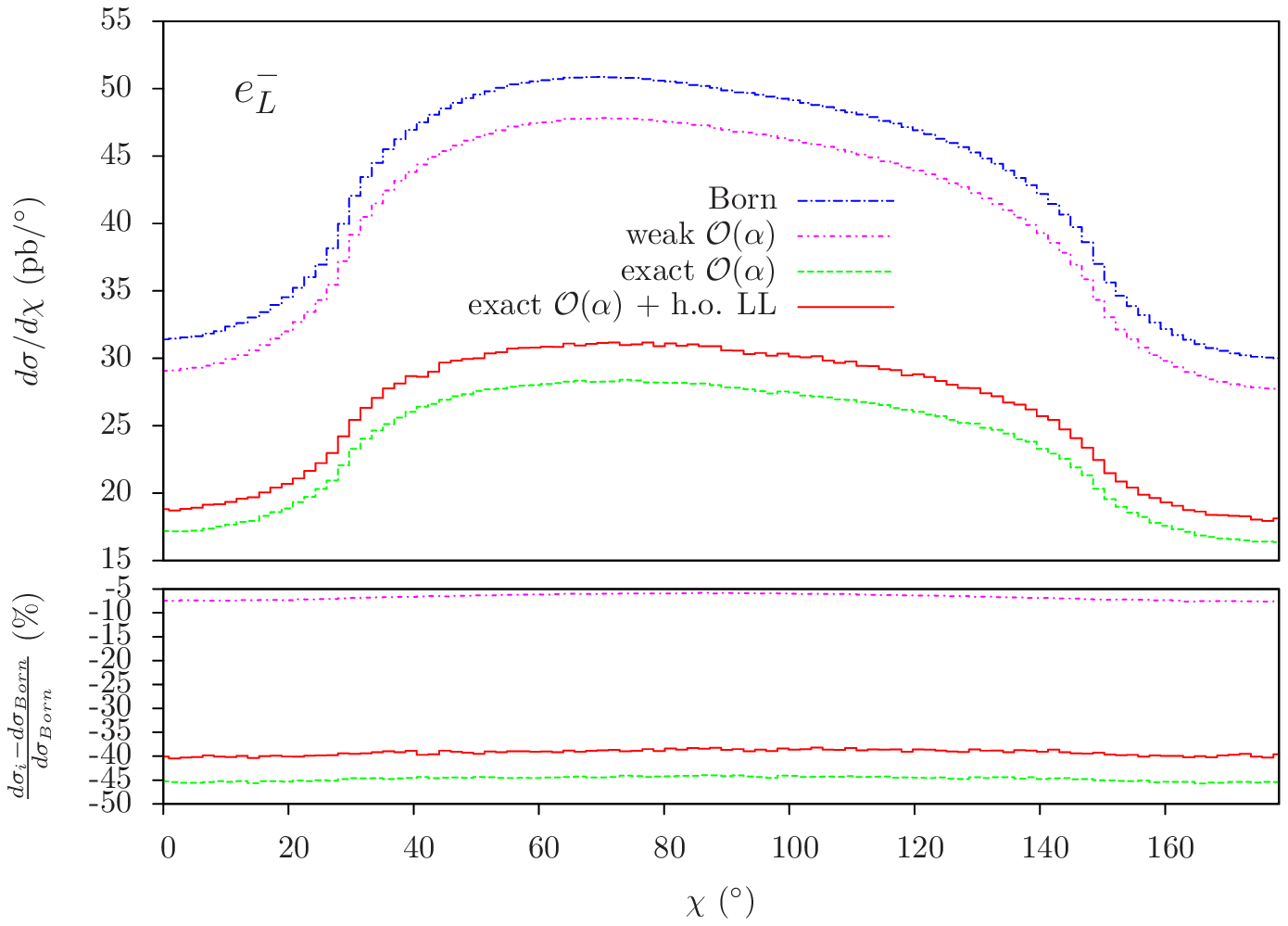}
\caption{Top/Bottom is for $e^-_R/e^-_L$: $\frac{d\sigma}{d\chi}$ distribution at the $Z$ peak.  (See the main text for the definition of this variable.)}
\label{chi-peak}}\end{center}\end{figure}

\begin{figure}\begin{center}{
\includegraphics[width=12.5cm]{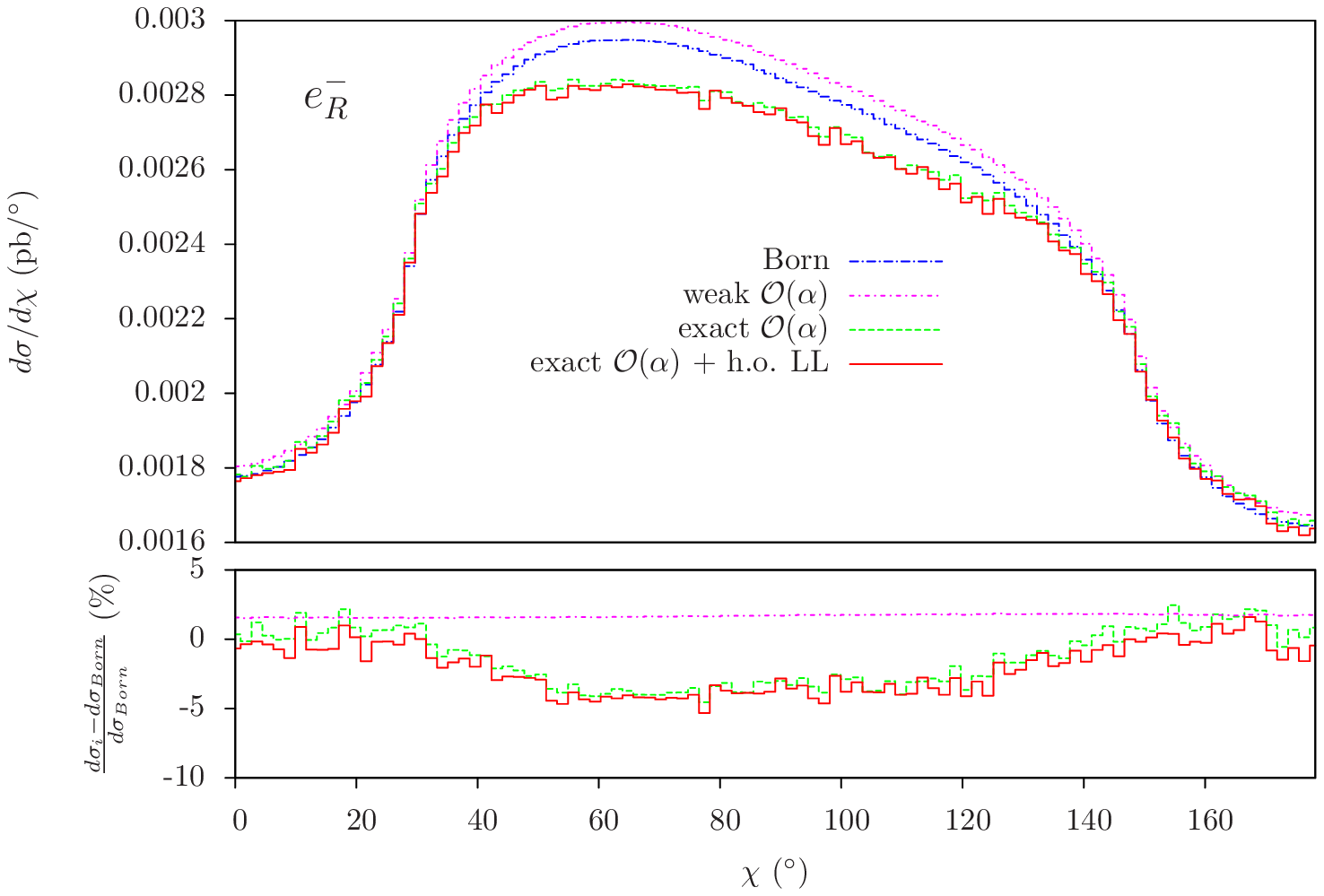}
\includegraphics[width=12.5cm]{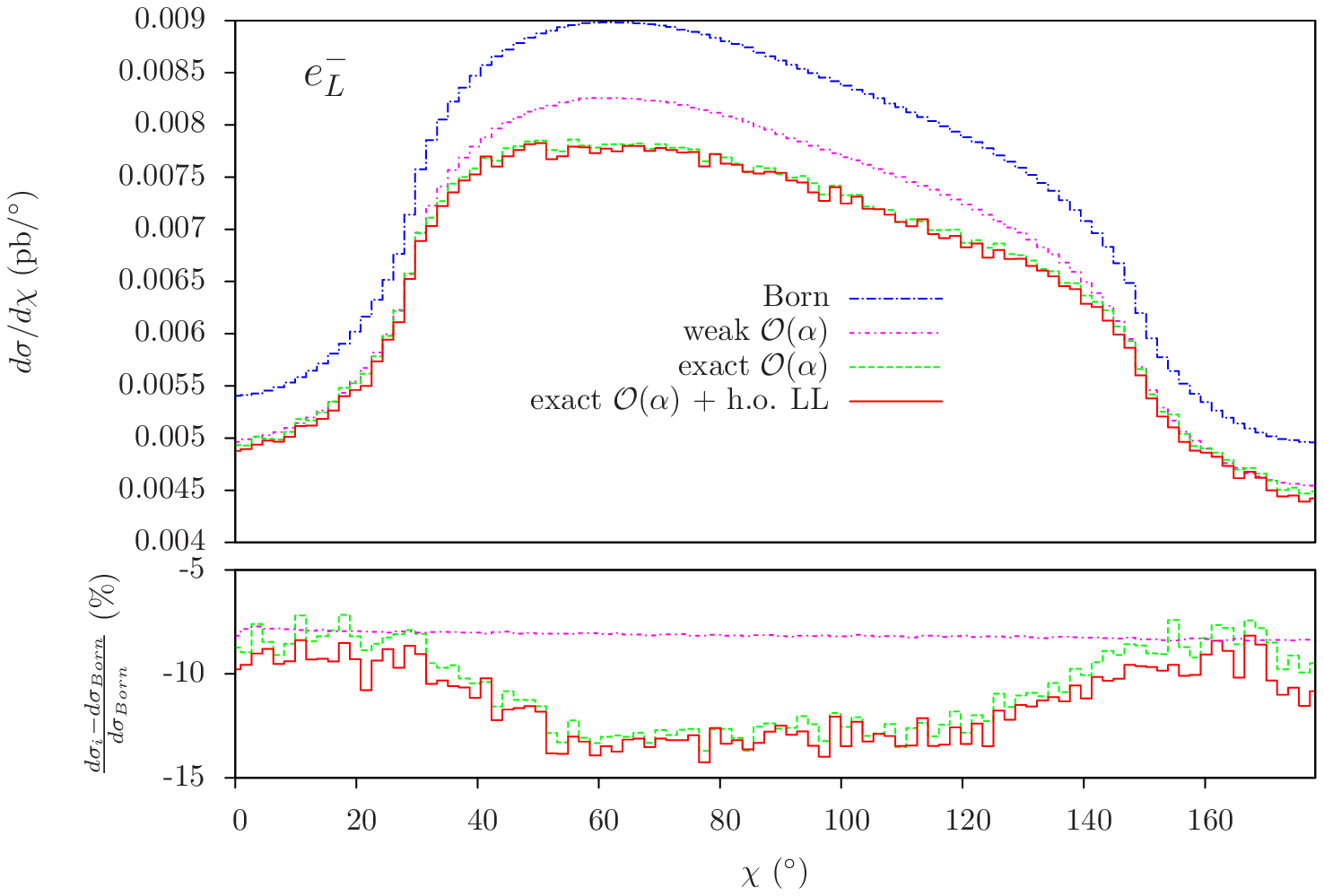}
\caption{Top/Bottom is for $e^-_R/e^-_L$: $\frac{d\sigma}{d\chi}$ distribution at 350 GeV.  (See the main text for the definition of this variable.)}
\label{chi-350}}\end{center}\end{figure}

\begin{figure}\begin{center}{
\includegraphics[width=12.5cm]{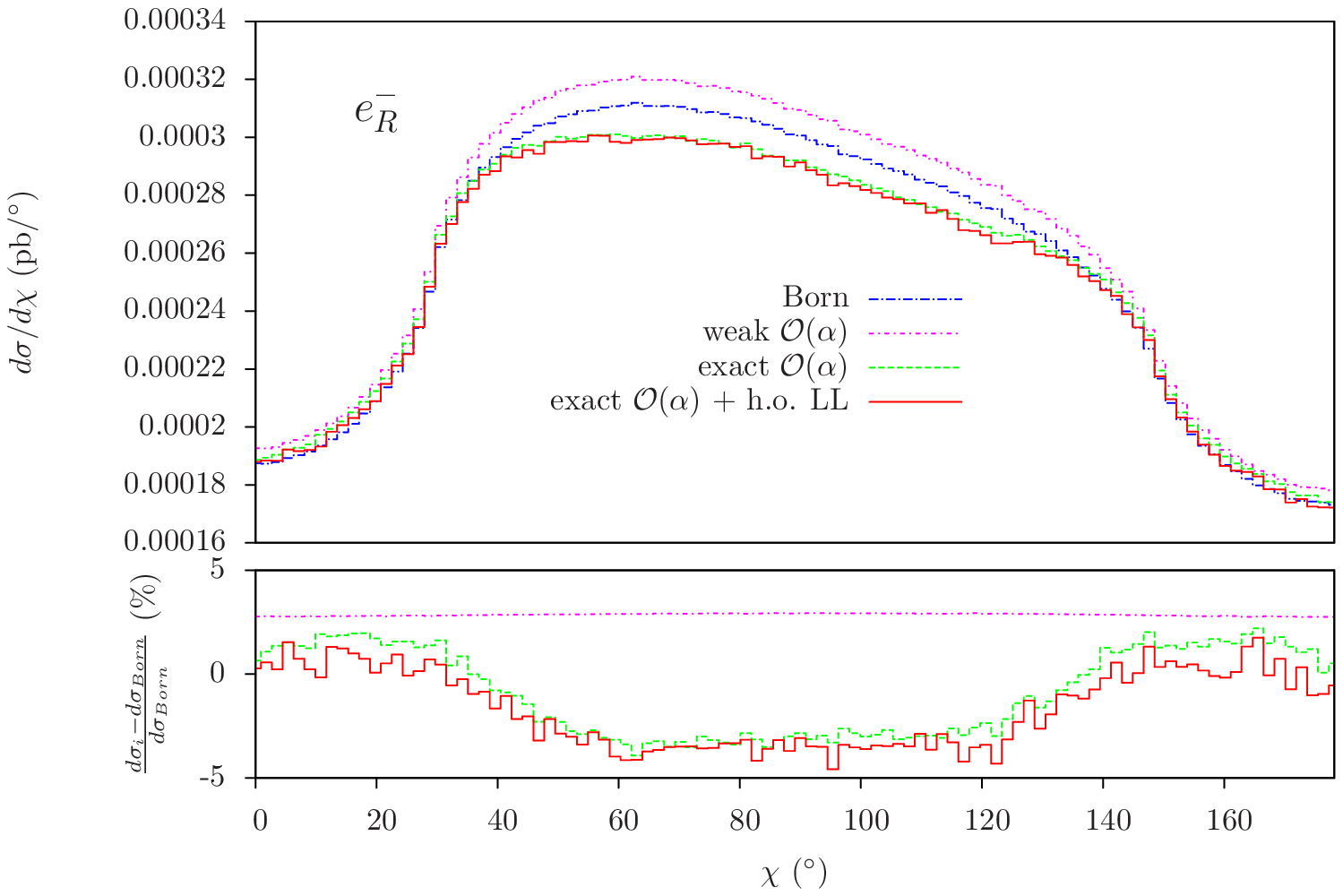}
\includegraphics[width=12.5cm]{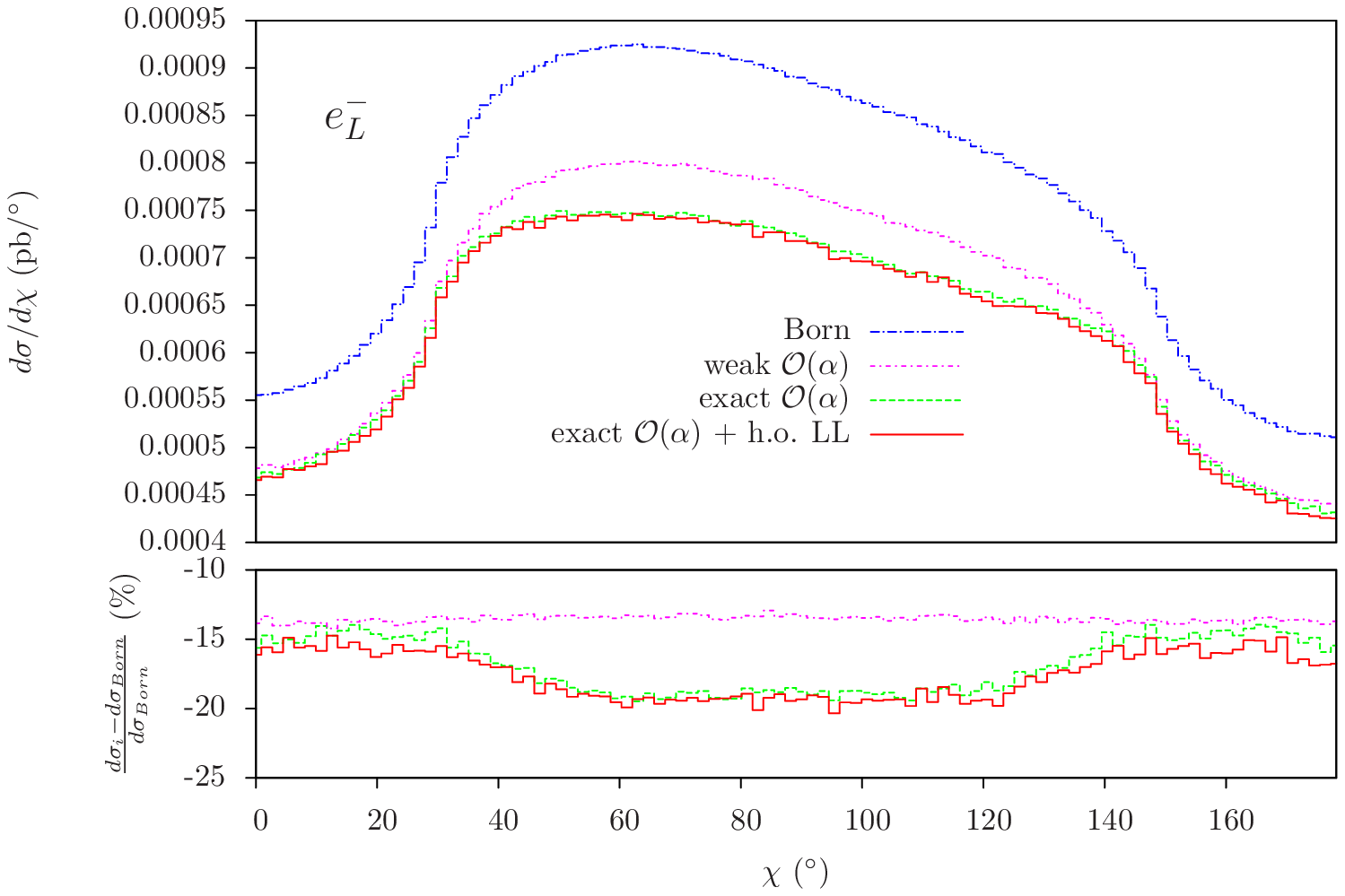}
\caption{Top/Bottom is for $e^-_R/e^-_L$: $\frac{d\sigma}{d\chi}$ distribution at 1 TeV.  (See the main text for the definition of this variable.)}
\label{chi-1000}}\end{center}\end{figure}

In the semi-inclusive case we found little differences with respect to 
the fully inclusive setup in the case of both the polar and azimuthal
angles, hence we will not dwell upon these variables here. This is
also true for the polar angle in the case of the exclusive analysis,
so that we do not treat this variable either. In contrast,  
the azimuthal
distribution, which we now define as  (again, with obvious meaning of the 
vector symbols)
\begin{equation}
\cos\chi = {\vec q \times \vec g \over{
| \vec q \times \vec g |}}
\cdot {\vec q \times \vec{e^-} \over {
| \vec q \times \vec{e^-}|}},
\label{chi}
\end{equation}
 is affected somewhat differently in the
exclusive case compared with the fully inclusive one.
This is displayed in Figs.~\ref{chi-peak}--\ref{chi-1000}.
Whilst at $\sqrt s=M_Z$ differences between these two cases are
really minimal and, in fact, mainly due to the shape of the tree-level
distributions, at  $\sqrt s=350$ GeV and 1 TeV, the 
${\cal O}(\alpha_{\rm{EM}})$ effects are different in both shape and
size. Whilst the QED ones still display a maximum 
(negative) correction at $\chi=90^{\rm o}$, they also show
local maxima (either positive or negative) 
in the near forward/backward directions,
for both energies and both electron polarisations. Sizewise, in the exclusive
case, QED corrections are smaller than in the fully inclusive case, with
this reduction being more pronounced (by a factor of 3 or so) for 
 right-handed  rather than left-handed 
(by a factor of 1.5 or so) electrons. (Also notice that  QED LL resummation
effects through higher orders are never relevant.)
Finally, the purely weak corrections in the exclusive case are never
substantially different from the counterparts in the fully inclusive scenario.

\section{Summary}
\label{Sec:Summary}

In summary, a careful analysis of actual $e^+e^-$ $\to$ $q\bar qg$ events,  involving one-loop
EW effects, is required, 
particularly in presence of beam polarisation and high energy, if one wants to accurately characterise the
event orientation. We have in fact found  substantial ${\cal O}(\alpha_{\rm{EM}})$
corrections in the case of both the QED and weak components that render the full one-loop
${\cal O}(\alpha_{\rm S}\alpha_{\rm{EM}}^3)$ predictions measureably different from  the tree-level
${\cal O}(\alpha_{\rm S}\alpha_{\rm{EM}}^2)$ ones for both polar and azimuthal angular distributions
used to identify the event orientation, irrespective of
 the level of inclusiveness (or exclusiveness) of the 
analysis. Hence, since such genuine SM effects could well mimick new physics phenomena, it
is of paramount importance that they are accounted for in phenomenological
studies of the three-jet event sample. In our calculation, we have neglected tree-level $W^\pm$ and $Z$ bremsstrahlung
through ${\cal O}(\alpha_{\rm S}\alpha_{\rm{EM}}^3)$, as we expect that,
in view of the cleanliness of three-jet samples produced in electron-positron machines  compared
with  hadronic ones, their contribution can be 
efficiently identified  and removed from the data.

\clearpage

\clearpage
\clearpage
\clearpage

\clearpage
\clearpage
\clearpage

\clearpage
\clearpage
\clearpage

\end{document}